\documentclass[10pt,pra,aps,twocolumn,superscriptaddress,floatfix,notitlepage,nofootinbib,reprint]{revtex4-2}
\usepackage{amsmath}
\usepackage{amsfonts}
\usepackage{amssymb}
\usepackage[pdftex]{graphicx}
\usepackage{hyperref}
\usepackage{xcolor}
\usepackage{physics}
\usepackage{subcaption}
\usepackage{comment}
\usepackage{dsfont}
\usepackage{braket}
\usepackage[inkscapelatex=false]{svg}
\usepackage{witharrows}
\usepackage{soul}

\begin{document}
\title{Nonclassicality of Mixed States with Photon Number Coherence}
\author{Spencer Rogers}
\email{spencer.rogers@uri.edu}
\affiliation{Department of Physics, University of Rhode Island, Kingston, RI 02881, USA}

\author{Salman Shahid}
\affiliation{Department of Physics, University of Rhode Island, Kingston, RI 02881, USA}

\author{Wenchao Ge}
\email{wenchao.ge@uri.edu}
\affiliation{Department of Physics, University of Rhode Island, Kingston, RI 02881, USA}

\date{\today}
\begin{abstract}
The operational resource theory (ORT) measure is a nonclassicality measure for bosonic states, notable for its resource-theoretic properties and connection to metrology. However, it can be difficult to evaluate, being linked to an optimization problem for mixed states. Here, we present ORT measure calculations for mixed states with photon number coherence. We give exact formulas governing the ORT measure of a broad class of rank-two mixed states, and numerical solutions for some higher-rank states. We also compare the nonclassicality of these states to their metrological power, thus showing in what regimes the metrological power manages to saturate the ORT bound. Throughout, we consider the role of coherence. In particular, we show that nonclassicality and metrological power never increase under bosonic dephasing, but may plateau in a manner similar to entanglement sudden death. Nevertheless, lowering photon number coherence more freely can sometimes yield more nonclassical and metrologically useful states.
\end{abstract}
\maketitle

\section{Introduction}

Nonclassicality, the ability of a quantum optical system to be in a superposition of classical (coherent) states, offers powerful advantages in metrology \cite{caves1981quantum,Lang13,giovannetti2004quantum,barnett2003ultimate,holland1993interferometric,campos2003optical,pezze2008machzehnder,birrittella2012multiphoton,Ge2023njp}. For example, squeezed vacuum states have been used to push gravitational-wave sensing beyond the standard limit, improving sensitivity despite limitations on laser power \cite{ligo2011gravitational,aasi2013enhanced,tse2019quantum}. 

The Glauber-Sudarshan $P$-function \cite{glauber63coherent,Sudarshan63} sets the fundamental criterion for nonclassicality, the basic idea being that a nonclassical state is one that cannot be represented as a positive probability distribution over coherent states. Consistent with this basic criterion, nonclassicality may be further quantified based on various standards, such as: minimum geometric distance to a classical state \cite{lee1991measure}, shortest coherent state expansion \cite{gehrke2012quantification,vogel2014unified}, entanglement potential \cite{asboth2005computable}, sub-Poissonianity  \cite{mandel1979sub}, and resource-theoretic considerations \cite{gehrke2012quantification,sperling2015convex,tan2017quantifying,YadinPRX18}.

Among the proposed nonclassicality measures, the operational resource theory (ORT) of nonclassicality \cite{ge2020operational,ge2020evaluating,merlin2022operational,rogers2024quantifying} is particularly well-suited for metrology, due to its resource-theoretic nature and connection to sensing applications (see Ref. \cite{Kwon19,YadinPRX18} for similar constructions). This theory treats nonclassicality as a resource that does not increase under certain free operations (dubbed ``classical operations"), such as linear optical operations. The ORT measure, which quantifies the amount of this resource in a given state, is related to the usefulness of a state for sensing tasks: equivalent to the metrological power of pure states, and an upper bound on the metrological power of mixed states. This is the closest possible relationship because some nonclassical mixed states do not offer any metrological advantage over classical states.

Like entanglement measures, the ORT measure is difficult to evaluate for mixed states, being linked to a convex roof optimization problem \cite{toth2015evaluating,rothlisberger2012libcreme,zhu2025unified}. Previous works provided solutions for incoherent mixtures of Fock states \cite{ge2020evaluating,rogers2024quantifying}, whose symmetry (i.e., invariance under phase shifts) simplifies the optimization problem. Solving for the ORT measure of mixed states without this symmetry may yield new insights, for example into how coherence, dephasing, and loss impact nonclassicality and metrological power.

This paper provides results highlighting, primarily, the subtle relationship between coherence and nonclassicality. We prove that the nonclassicality of a state without photon number coherence is never greater than that of a partially-coherent state with the same photon number populations. We also introduce a class of rank-two states whose properties permit a precise analytical solution for the ORT measure across varying values of population and coherence. While limited, this class includes relevant examples wherein the impact of dephasing, and in some cases, loss, on nonclassicality can be quantified. Within this class, we witness a general trend of piecewise behavior: for example, reducing coherence via a continuous dephasing process may at first reduce nonclassicality, but stop doing so once a certain threshold is reached, similar to ``entanglement sudden death" \cite{yu2004finite,yu2009sudden,zyczkowski2001dynamics,Diosi2003,dodd2004disentanglement}. We supplement these results with metrological power calculations, and some higher-rank examples. One of these shows that reducing number coherence can sometimes yield more nonclassical and metrologically useful states, if this reduction is inconsistent with bosonic dephasing.

Our paper is structured as follows. In section II, we review the ORT measure and show how its convexity property may be used to compare the nonclassicalities of states with different photon number coherence. In section III, we present exact, general formulas that govern the  ORT measure and metrological power of a broad class of rank-two mixed states. Section IV extends this class of states by adding off-diagonal elements, i.e., coherences. Section V gives numerical solutions for higher-rank states.


\section{The ORT Measure and its Properties}
In this section, we introduce the operational resource theory measure of nonclassicality and relate its definition to a nonclassicality witness and another nonclassicality measure. We then discuss the basic properties within a quantum resource theory and its operational significance for quantum metrology. As a preliminary, we also explore the effect of photon-number coherence on the ORT measure qualitatively. 
\subsection{The ORT Measure}
Our system of interest is a single optical mode. Equivalently, one may think of this as a quantum harmonic oscillator, also known as a bosonic system. A \textit{pure} bosonic state $\ket{\psi}$ is regarded as \textit{classical} if and only if $\hat{a}\ket{\psi}=\alpha\ket{\psi}$~\cite{Hillery:1985aa}, for some complex number $\alpha$ and the annihilation operator $\hat{a}$. A pure state that does not satisfy the above relationship for any $\alpha$ is considered a \textit{nonclassical} state. Some examples are Fock states $\ket{n}$ (for $n>0$), cat states (which are discrete superpositions of coherent states)\ and squeezed vacuum. Given the applications of nonclassicality ~\cite{braunstein1998teleportation,braunstein2005quantum,mirrahimi2014dynamically,giovannetti2006quantum}, one may wish to quantify how nonclassical these different states are. 


A \textit{nonclassicality witness} $\mathcal{W}(\ket{\psi})\geq0$ is a function such that $\mathcal{W}(\ket{\psi})=0$ for all classical states. A value $\mathcal{W}>0$ thus implies a nonclassical state. An example of a nonclassicality witness is $\mathcal{W}(\ket{\psi})=|\langle\hat{a}^2\rangle-\langle\hat{a}\rangle^2|$, which quantifies the squeezing of a state \cite{ge2020operational}. $\mathcal{W}(\ket{\alpha})=|\alpha^2-\alpha^2|=0$ for all coherent states $\alpha$, while $\mathcal{W}(\ket{\psi})>0$ for some nonclassical states: e.g., for a squeezed vacuum state $\mathcal{W}(\ket{\text{SV}})=\sinh\gamma\cosh\gamma$, where $\gamma$ is the squeezing strength. Note that a witness may be zero for some nonclassical states: in our example, $\mathcal{W}(\ket{n})=0$ for all $\ket{n}$. 

A \textit{nonclassicality measure} $\mathcal{N}(\ket{\psi})\geq0$ is a function such that $\mathcal{N}(\ket{\psi})=0$ for all classical states and $\mathcal{N}(\ket{\psi})>0$ for all nonclassical states. A value $\mathcal{N}>0$ implies a nonclassical state, while a value $\mathcal{N}=0$ implies a classical state. An example of a nonclassicality measure is $\mathcal{Q}(\ket{\psi})=\langle\hat{a}^\dagger\hat{a}\rangle-|\langle\hat{a}\rangle|^2$~\cite{Kwon19}, which represents the energy that cannot be removed via displacement operations \cite{ge2020operational}.

If one takes a nonclassicality measure $\mathcal{N}$, and adds a nonclassicality witness $\mathcal{W}$ to it, the sum $\mathcal{N}+\mathcal{W}$ is another nonclassicality measure. Using the measure and witness from the above, we obtain the measure:

\begin{equation}
\mathcal{N}_{\text{ORT}}(\ket{\psi})=\langle\hat{a}^\dagger\hat{a}\rangle-|\langle\hat{a}\rangle|^2+|\langle\hat{a}^2\rangle-\langle\hat{a}\rangle^2|.
\label{eq:pureStateORT}
\end{equation}
This is the operational resource theory measure (ORT) of nonclassicality for pure states~\cite{ge2020operational, YadinPRX18}. It is \textit{operational} in that it relates to the metrological power of quadrature variance: $\mathcal{N}_{\text{ORT}}=\max_\mu\langle(\Delta\hat{X}_\mu)^2\rangle-1/2$, where $\hat{X}_\mu=i(e^{-i\mu}\hat{a}^\dagger-e^{i\mu}\hat{a})/\sqrt{2}$ is a quadrature. A larger value indicates greater susceptibility to displacements in the conjugate variable $\hat{P}_\mu$, which may be exploited for sensing. Note that such nonclassical displacement-sensitivity may be converted to enhanced phase-sensitivity in a multi-mode linear optical scheme employing a strong classical source \cite{Ge2023njp}. $\mathcal{N}_{\text{ORT}}$ is \textit{resource-theoretic} in that it does not increase under certain free operations (like displacements and phase shifts). Henceforth, we will drop the subscript $_{\text{ORT}}$ and refer to $\mathcal{N}_{\text{ORT}}$ as $\mathcal{N}$; similarly, when discussing ``nonclassicality" quantitatively, we will generally mean the ORT measure of nonclassicality.

The notions of classicality and nonclassicality may be extended to mixed states as well. A mixed state $\hat{\rho}$ is regarded as \textit{classical} if and only if it is equivalent to an ensemble of coherent states: $\hat{\rho}=\sum_\alpha q_\alpha\ket{\alpha}\bra{\alpha}$ where $q_\alpha>0$ (the summation may be continuous or discrete) ~\cite{lee1991measure,SZ}. The ORT measure for mixed states is $\mathcal{N}(\hat{\rho})=\min_{\{q_j,\ket{\phi_j}\}}\left[\max_\mu\left(\sum_jq_j\bra{\phi_j}(\Delta \hat{X}_\mu)^2\ket{\phi_j}\right)-1/2\right]$, or equivalently~\cite{ge2020operational}:
\begin{widetext}
\begin{equation}
\mathcal{N}(\hat{\rho})=\min_{\{q_j,\ket{\phi_j}\}}\left(\underbrace{\langle{\hat{a}^\dagger\hat{a}}\rangle-\sum_j q_j\left|\bra{\phi_j}\hat{a}\ket{\phi_j}\right|^2}_{\text{``measure" term}}+\underbrace{\left|\langle{\hat{a}^2}\rangle-\sum_j q_j\bra{\phi_j}\hat{a}\ket{\phi_j}^2\right|}_{\text{``witness" term}}\right).
\label{eq:ORTMixed}
\end{equation}
\end{widetext}
The minimization is taken over all possible decompositions $\{q_j,\ket{\phi_j}\}$ of $\hat{\rho}$, satisfying $\hat{\rho}=q_j\ket{\phi_j}\bra{\phi_j}$ with $q_j>0$ for all $q_j$. The optimal decomposition is generally not unique (see Appendix B of Ref. \cite{rogers2024quantifying}). We call one term in Eq.~\eqref{eq:ORTMixed} the ``measure term" because:
\begin{equation}
\mathcal{Q}(\hat{\rho})=\min_{\{q_j,\ket{\phi_j}\}}\left(\langle{\hat{a}^\dagger\hat{a}}\rangle-\sum_j q_j\left|\bra{\phi_j}\hat{a}\ket{\phi_j}\right|^2\right)
\label{eq:measureTerm}
\end{equation} is itself a nonclassicality measure~\cite{Kwon19}. We call the other term the ``witness term" because:
\begin{equation}
\mathcal{W}(\hat{\rho})=\min_{\{q_j,\ket{\phi_j}\}}\left|\langle{\hat{a}^2}\rangle-\sum_j q_j\bra{\phi_j}\hat{a}\ket{\phi_j}^2\right|
\label{eq:witnessTerm}
\end{equation}
is a nonclassicality witness. Note that, $\mathcal{N}(\hat{\rho})=\mathcal{Q}(\hat{\rho})+\mathcal{W}(\hat{\rho})$ if and only if $\mathcal{Q}$ and $\mathcal{W}$ have a common optimal decomposition, otherwise $\mathcal{N}>\mathcal{Q}+\mathcal{W}$.

The ORT measure for mixed states, Eq.~\eqref{eq:ORTMixed}, has several nice properties:

(i) Non-negativity: $\mathcal{N}(\hat{\rho})\geq0$, where equality holds if and only if $\hat{\rho}$ is classical. This is the minimum requirement of a nonclassicality measure.

(ii) Weak monotonicity: $\mathcal{N}(\Lambda[\hat{\rho}])\leq\mathcal{N}(\hat{\rho})$ for classical operations $\Lambda$. Mathematically, $\Lambda(\rho)=\Tr_A[\hat{U}\hat{\rho}\otimes\hat{\rho}_A\hat{U}^\dagger]$, where $\hat{\rho}_A$ is a classical ancillary state and $\hat{U}$ is a linear optical unitary \cite{ge2020operational}. Weak monotonicity is the resource-theoretic property of the measure.

(iii) Convexity: $\sum_jp_j\mathcal{N}(\rho_j)\geq\mathcal{N}(\sum_jp_j\hat{\rho}_j)$.

(iv) Lower-bounded by metrological power (of quadrature variance): $\mathcal{N}(\hat{\rho})\geq\mathcal{M(\hat{\rho})}$, where equality holds for pure states. The metrological power is defined as 
\begin{equation}
\label{eq:MetroPower}
    \mathcal{M}(\hat{\rho})\equiv\max[F_X(\hat{\rho})-1/2,0],
\end{equation}
where\newline
$F_X(\hat{\rho})=\max_\mu\left[\min_{\{q_j,\ket{\phi_j}\}}\left(\sum_jq_j\bra{\phi_j}(\Delta \hat{X}_\mu)^2\ket{\phi_j}\right)\right]$
is one fourth of the Quantum Fisher Information for the optimal quadrature~\cite{ge2020operational}. Given the eigen-decomposition of $\hat{\rho}=\Sigma_k \lambda_j\ket{\lambda_j}\bra{\lambda_j}$, we have~\cite{ge2020operational}:
\begin{equation}    F_X(\hat{\rho})=\max_{\mu}\left[\text{Tr}\left[\hat{X}_\mu^2\hat{\rho}\right]-\sum_{j,k}\left|\bra{\lambda_j}\hat{X}_\mu\ket{\lambda_k}\right|^2\frac{2\lambda_j\lambda_k}{\lambda_j+\lambda_k}\right].\label{eq:fisherEigendecomposition}
\end{equation}
The metrological power $\mathcal{M}$ describes the sensing enhancement over classical states: for all classical states, $\mathcal{M}=0$. Notably, there exist nonclassical mixed states for which $\mathcal{M}=0$, meaning $\mathcal{M}$ is only a nonclassicality witness, as opposed to a measure ~\cite{ge2020operational,ge2020evaluating}. Thus, the tight inequality $\mathcal{N}(\hat{\rho})\geq\mathcal{M}(\hat{\rho})$ is the closest possible relationship between a nonclassicality measure and the metrological power.

Calculating the ORT measure from Eq.~\eqref{eq:ORTMixed} is a nontrivial optimization problem. It is similar to convex roof optimization problems that arise in entanglement quantification --- an entanglement measure for mixed states often takes the form of the minimum average over pure states: $\mathcal{E}(\hat{\rho})=\min \sum_jq_j\mathcal{E}(\ket{\phi_j})$ \cite{chen2005entanglement}. Semidefinite programming, and gradient methods over Stiefel manifolds, have been used to solve for the minimum in these contexts \cite{toth2015evaluating,rothlisberger2012libcreme,zhu2025unified}. However, in general the ORT measure $\mathcal{N}(\hat{\rho})\neq\min \sum_jq_j\mathcal{N}(\ket{\phi_j})$, unlike the measure introduced in Ref.~\cite{YadinPRX18}, owing to the absolute value in the witness term. 

Previously, it was shown that if the state $\hat{\rho}$ in question is an incoherent mixture of Fock states, $\hat{\rho}_{F}=\sum_{n=0}^{M-1}p_n\ket{n}\bra{n}$, the witness term can be ignored without loss of generality and $\mathcal{N}(\hat{\rho}_F)=\mathcal{Q}(\hat{\rho}_F)=\min_j\sum_jq_j\mathcal{Q}(\ket{\phi}_j)$. Linear programming was then used to evaluate the measure (semidefinite programming and gradient methods are also valid) \cite{rogers2024quantifying}. However, this left out a variety of cases with coherence between Fock states, i.e., photon number coherence. This work presents ORT measure calculations for mixed states with photon number coherence.

\subsection{Effect of Photon Number Coherence}\label{sec:EffectOfPhotonNumberCoherence}

Before discussing our main results, let us consider qualitatively how photon number coherence, i.e, the magnitude of off-diagonal elements $\rho_{nm}\equiv\bra{n}\hat{\rho}\ket{m}$, should affect a mixed state's nonclassicality. For this, it is helpful to introduce a bosonic dephasing channel $\mathcal{D}_p(\hat{\rho})$, which acts as:
\begin{equation}
\mathcal{D}_p(\hat{\rho})\equiv\int_{-\pi}^\pi d\phi \hspace{0.1cm} p(\phi)e^{-i\hat{n}\phi}\hat{\rho}e^{i\hat{n}\phi}
\end{equation}
for some characteristic probability density function $p(\phi)$ \cite{huang2024exact}. A bosonic dephasing channel may be thought of as a statistical mixture of phase shifts. Note that a definite phase shift has no effect on the ORT measure: $\mathcal{N}(e^{-i\hat{n}\phi}\hat{\rho}e^{i\hat{n}\phi})=\mathcal{N}(\hat{\rho})$ (see Appendix \ref{sec:appendixPhaseShift}). Combining this with the convexity (iii) property of the measure, we obtain:
\begin{equation}
\mathcal{N}(\mathcal{D}_p(\hat{\rho}))\leq\mathcal{N}(\hat{\rho}).
\label{eq:generalBDCBound}
\end{equation}
This is important, because while bosonic dephasing channels have no effect on photon number populations, they can lower photon number coherences. Since metrological power $\mathcal{M}$ also does not change under a definite phase shift (see Appendix \ref{sec:appendixPhaseShift}), and satisfies convexity \cite{Kwon19}, it satisfies the same relation:
\begin{equation}
\mathcal{M}(\mathcal{D}_p(\hat{\rho}))\leq\mathcal{M}(\hat{\rho}).
\label{eq:metrologicalPowerBDCBound}
\end{equation}

By choosing a uniformly uncertain phase shift, $p(\phi)=(2\pi)^{-1}$, one obtains the ``total dephasing channel" $\mathcal{D}_\text{tot}$ that zeroes out all photon number coherences of any input state:
\begin{equation}
\mathcal{D}_\text{tot}\left(\sum_np_{n}\ket{n}\bra{n}+\sum_{n\neq m}\rho_{nm}\ket{n}\bra{m}\right)=\sum_np_{n}\ket{n}\bra{n}.
\label{eq:totalDephasingChannel}
\end{equation}
Combining Eq.~\eqref{eq:generalBDCBound} and~\eqref{eq:totalDephasingChannel}, we learn that:
\begin{equation}
\begin{split}
&\mathcal{N}\left(\sum_np_{n}\ket{n}\bra{n}\right)\leq\\
&\mathcal{N}\left(\sum_np_{n}\ket{n}\bra{n}+\sum_{n\neq m}\rho_{nm}\ket{n}\bra{m}\right).
\end{split}
\label{eq:noCoherenceVsWithCoherence}
\end{equation}
This provides significant context to our paper: adding photon number coherence to incoherent mixtures of Fock states can only increase nonclassicality (or leave it the same). An analogous relation holds for metrological power $\mathcal{M}$.

While Eq.~\eqref{eq:noCoherenceVsWithCoherence} provides a neat comparison of the nonclassicalities of states with photon number coherence against those completely without it, more generally we would like to compare the nonclassicality of a state with photon number coherence against a state with ``less" photon number coherence. In some cases, symmetric dephasing channels:
\begin{equation}
\mathcal{D}_{\text{sym},p}(\hat{\rho})\equiv\int_{0}^\pi d\phi \hspace{0.1cm} \frac{p(\phi)}{2}\left(e^{-i\hat{n}\phi}\hat{\rho}e^{i\hat{n}\phi}+e^{i\hat{n}\phi}\hat{\rho}e^{-i\hat{n}\phi}\right),
\label{eq:symmetricDephasing}
\end{equation}
provide this comparison. A symmetric dephasing channel has the effect:
\begin{equation}
\begin{split}
&\mathcal{D}_{\text{sym},p}\left(\sum_np_{n}\ket{n}\bra{n}+\sum_{n\neq m}\rho_{nm}\ket{n}\bra{m}\right)\\
&=\sum_np_{n}\ket{n}\bra{n}+\sum_{n\neq m}\rho_{nm}\langle\cos((n-m)\phi)\rangle\ket{n}\bra{m},
\label{eq:symmetricDephasingChannelEffect}
\end{split}
\end{equation}
where $\langle\cos((n-m)\phi)\rangle=\int_0^\pi d\phi \hspace{0.1cm} p(\phi)\cos((n-m)\phi)$. Thus, symmetric dephasing channels may reduce the magnitude of photon number coherences (possibly changing their sign as well). Of course, from Eq.~\eqref{eq:generalBDCBound}, $\mathcal{N}(\mathcal{D}_{\text{sym},p}(\hat{\rho}))\leq\mathcal{N}(\hat{\rho})$. It is important not to draw too-strong conclusions from  this, however. Given a pair of states $\hat{\rho}$ and $\hat{\rho}'$ with the same photon number populations, there may not be any bosonic dephasing channel connecting them, even if $\hat{\rho}'$ has strictly smaller off-diagonal elements $\rho'_{nm}\leq\rho_{nm}$! Bosonic dephasing cannot decrease photon number coherences arbitrarily. In Appendix \ref{sec:appendixDephasing}, we illustrate these limitations for partially-coherent mixtures of three Fock states.

\begin{figure}
  \centering
   \captionsetup{justification=raggedright, singlelinecheck=false}
  \includegraphics[width=\columnwidth]{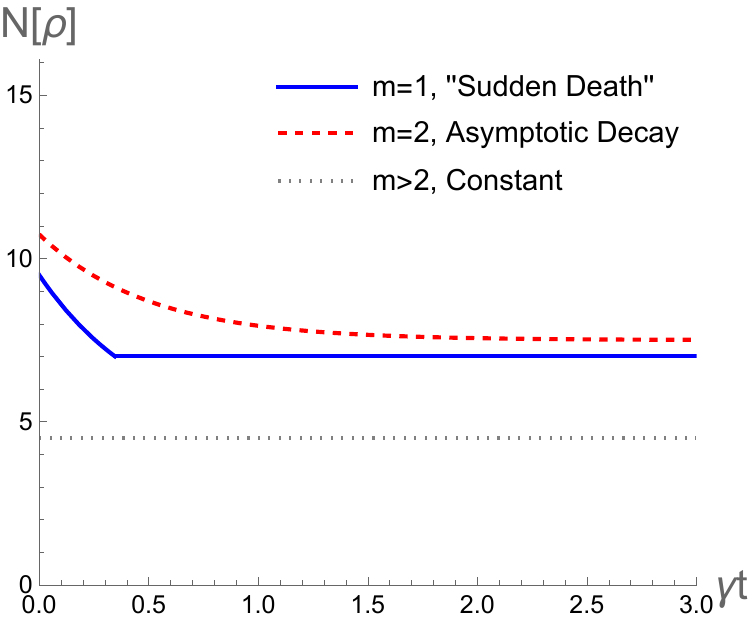}
\caption{Effect of dephasing $\ket{n+m}\bra{n}\rightarrow e^{-im\omega_0t}e^{-|m|\gamma t}\ket{n+m}\bra{n}$ (as due to a Lorentzian frequency profile) on the ORT measure $\mathcal{N}(\hat{\rho})$ for states that are, at $t=0$, pure superpositions $\sqrt{p_{n+m}}\ket{n+m}+\sqrt{1-p_{n+m}}e^{i\chi}\ket{n}$. Particular values for the $m=1$ curve are: $n=9$, $p_{10}=0.5$. For the $m=2$ curve: $n=6$, $p_8=0.75$. For the $m>2$ curve: $m=3$, $n=3$, and $p_6=0.5$.}
\label{fig:dephasing}
\end{figure}

The situation becomes much simpler if $\hat{\rho}$ and $\hat{\rho}'$ are partially-coherent mixtures of just two photon numbers $\ket{n+m}$ and $\ket{n}$, with the exact same photon number populations $p_{n+m}'=p_{n+m}$ and $p_{n}'=p_n$. Then, they are \textit{always} connected by a bosonic dephasing channel: if $\left|\rho'_{n+m,n}\right|<\left|\rho_{n+m,n}\right|$, then $\mathcal{N}(\hat{\rho}')\leq\mathcal{N}(\hat{\rho})$. When $m>2$, $\mathcal{N}(\hat{\rho})=\langle\hat{n}\rangle$, and does not depend on the photon number coherence at all (the bound in Eq.~\eqref{eq:noCoherenceVsWithCoherence} is always saturated). When $m=2$, $\mathcal{N}(\hat{\rho})=\langle\hat{n}\rangle+\left|\langle\hat{a}^2\rangle\right|=\langle\hat{n}\rangle+\left|\bra{n+2}\hat{\rho}\ket{n}\right|\sqrt{(n+2)(n+1)}$. There is a contribution that scales linearly with the photon number coherence.


When $m=1$, we have, from previous works \cite{ge2020evaluating,rogers2024quantifying}, the ORT measure of the pure superposition and the completely incoherent mixture:
\begin{subequations}
\begin{align}
&\mathcal{N}(\sqrt{p}\ket{n+1}+e^{i\chi}\sqrt{1-p}\ket{n})=\langle\hat{n}\rangle\\
\begin{split}
&\mathcal{N}\left(p\ket{n+1}\bra{n+1}+(1-p)\ket{n}\bra{n}\right)\\
&=\langle\hat{n}\rangle -p(1-p)(n+1).
\end{split}
\end{align}    
\end{subequations}
Our new results (particularly those in section \ref{sec:addingCoherence}) show how the ORT measure transitions between the two values in a dephasing process, see Fig. \ref{fig:dephasing}. Interestingly, once the photon number coherence sufficiently drops, the ORT measure reaches its final value, a situation similar to entanglement sudden death \cite{yu2004finite,yu2009sudden,zyczkowski2001dynamics,Diosi2003,dodd2004disentanglement}. Unlike in entanglement sudden death, however, nonclassicality does not reach a final value of zero. Instead, the plateau corresponds to a nonclassicality witness reaching zero. Similar vanishing of nonclassical effects was reported in Ref. \cite{bartkowiak2011sudden}.

\section{Analytical Solution for Class of Rank-2 Mixtures}\label{sec:Rank2NoCoherence}
We now present a class of rank-2 mixed states, whose ORT measure can be evaluated exactly. This class includes incoherent mixtures of Fock states, but also other mixed states with appropriate eigenstates. Section \ref{sec:addingCoherence} extends this class by introducing off-diagonal elements (thus addressing the partially-coherent mixture of Fock states alluded to at the end of the previous section).

A general rank-2 mixture, written in terms of its eigendecomposition, is:
\begin{equation}
\hat{\rho}=p\ket{\psi_1}\bra{\psi_1}+(1-p)\ket{\psi_2}\bra{\psi_2}
\end{equation}
where $\braket{\psi_2|\psi_1}=0$.\footnote{Note that, equivalently, $\hat{\rho}=p\ket{\psi_1'}\bra{\psi_1'}+(1-p)\ket{\psi_2'}\bra{\psi_2'}$, where the $\ket{\psi_j'}=e^{i\delta_j}\ket{\psi_j}$ are the same eigenstates but with different phases. The choice of phases $\delta_j$ may have an effect on the arguments of off-diagonal matrix elements: e.g., $\arg(\bra{\psi_j}\hat{a}\ket{\psi_k})=\arg(\bra{\psi_j'}\hat{a}\ket{\psi_k'})+\delta_j-\delta_k$.} Our analytically calculable rank-2 mixtures require two further assumptions. 


\textbf{Assumption 1: The two eigenstates are centered at the origin in phase space.} By this we mean that $\bra{\psi_1}\hat{a}\ket{\psi_1}=\bra{\psi_2}\hat{a}\ket{\psi_2}=0$. While the property $\bra{\psi}\hat{a}\ket{\psi}=0$ does not hold for general pure states, it does hold for certain well-known nonclassical states, including Fock states, even and odd cat states, and squeezed vacuum states.

\textbf{Assumption 2: The eigenstates satisfy $\arg(\bra{\psi_1}\hat{a}^2\ket{\psi_1})=\arg(\bra{\psi_2}\hat{a}^2\ket{\psi_2})=\left[\arg(\bra{\psi_1}\hat{a}\ket{\psi_2})+\arg(\bra{\psi_2}\hat{a}\ket{\psi_1})\right] \mod 2\pi$.} We consider any instance of $\arg(0)$ to be its own free variable, and we consider the equations satisfied if some choice yields equality. For example, this assumption is considered satisfied if $\bra{\psi_1}\hat{a}^2\ket{\psi_1}=\bra{\psi_1}\hat{a}\ket{\psi_2}=0$.

A standard scenario, that generates mixed states satisfying Assumptions 1 and 2, is subjecting a definite-parity pure state $\ket{\psi_0}=\sum_{n=0}^\infty c_ne^{i(2n+b)\phi}\ket{2n+b}$ (with $b\in\{0,1\}$ and $c_n\geq0$) to weak photon loss (we justify this in Appendix \ref{sec:appendixPhotonLossChannel}). To be clear, though, such states are only a subset, and our formulae apply to all rank-two states whose eigenstates satisfy Assumptions 1 and 2. Defining $r_{12}\equiv|\bra{\psi_1}\hat{a}\ket{\psi_2}|$, and $r_{21}\equiv|\bra{\psi_2}\hat{a}\ket{\psi_1}|$, we find the analytical expression of the ORT measure to be:
\begin{widetext}
\begin{DispWithArrows}<\mathcal{N}(\hat{\rho})=>[format = ll,subequations]
\langle\hat{n}\rangle+|\langle\hat{a}^2\rangle|-2p(1-p)(r_{21}+r_{12})^2, & \qquad  |\langle\hat{a}^2\rangle|\geq p(1-p)(r_{21}+r_{12})^2, \label{eq:overSqueezed} \\
\langle\hat{n}\rangle+\frac{-1}{r_{21}^2+r_{12}^2}\left[2r_{21}r_{12}|\langle\hat{a}^2\rangle|+(r_{21}^2-r_{12}^2)^2p(1-p)\right], &\qquad  |\langle\hat{a}^2\rangle|< p(1-p)(r_{21}+r_{12})^2.\label{eq:underSqueezed}
\end{DispWithArrows}
\end{widetext}
Details on the derivation may be found in Appendix \ref{sec:appendixDerivation}.

\begin{figure}
  \centering
   \captionsetup{justification=raggedright, singlelinecheck=false}
  \includegraphics[width=\columnwidth]{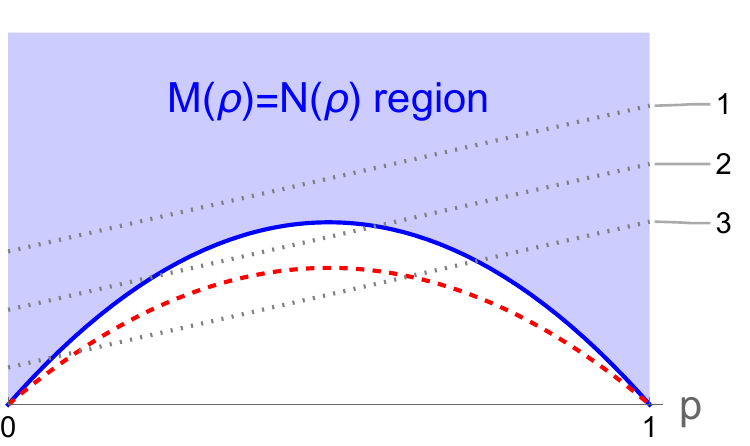}
\caption{Illustration of the three scenarios that can arise from the inequalities controlling $\mathcal{N}$ and $\mathcal{M}$ in section \ref{sec:Rank2NoCoherence}. The solid (blue) parabola represents $p(1-p)(r_{12}+r_{21})^2$, the dashed (red) parabola  represents $4p(1-p)r_{12}r_{21}$, and the dotted (gray) lines represent possible behaviors for $|\braket{\hat{a}^2}| =p|\braket{\hat{a}^2}_{1}|+(1-p)|\braket{\hat{a}^2}_{2}|$. Intersections of the dotted-gray line with the solid-blue (dashed-red) mark transitions between Eq. \eqref{eq:overSqueezed} and \eqref{eq:underSqueezed} for $\mathcal{N}$ (Eq. \eqref{eq:M1} and \eqref{eq:M2} for $\mathcal{M}$). For $p$ in which $|\langle\hat{a}^2\rangle|$ is above (below) the solid parabola, $\mathcal{M}=\mathcal{N}$ ($\mathcal{M}\leq\mathcal{N}$). Note that the particular properties of the lines and parabolas depend on eigenstate parameters $\{r_{12},r_{21},|\braket{\hat{a}^2}_{1}|,|\braket{\hat{a}^2}_{2}|\}$, and the particular values used to generate this figure are for illustration only. However, the solid (blue) parabola cannot be below the dashed (red) one. When $r_{21}=r_{12}$, both parabolas coincide and $\mathcal{M}=\mathcal{N}$ for all values of $p$. Note that, if $r_{12}=r_{21},$ the solid (blue) and dashed (red) curves coincide.}
  \label{fig:5cases}
\end{figure}

Assumption 2 implies that $|\langle\hat{a}^2\rangle|=p|\langle\hat{a}^2\rangle_1|+(1-p)|\langle\hat{a}^2\rangle_2|$ is first order in $p$. $\langle\hat{n}\rangle=p\langle\hat{n}\rangle_1+(1-p)\langle\hat{n}\rangle_2$ is also first-order in $p$. Thus, for fixed eigenbasis $\{\ket{\psi_1},\ket{\psi_2}\}$, $\mathcal{N}$ behaves as the piecewise concatenation of two quadratics in $p$ (Eq.~\eqref{eq:overSqueezed} and~\eqref{eq:underSqueezed}). The particular quadratic $\mathcal{N}$ follows (for each $p$) depends on whether the line $p|\langle\hat{a}^2\rangle_1|+(1-p)|\langle\hat{a}^2\rangle_2|$ is above the quadratic $p(1-p)(r_{21}+r_{12})^2$ (see Fig.~\ref{fig:5cases}). It follows that Eq.~\eqref{eq:underSqueezed} can hold for, at most, one finite $p$-interval, $(p_L,p_R)$.

Eq.~\eqref{eq:overSqueezed} corresponds to a nonzero nonclassicality witness, $\mathcal{W}$ in Eq.~\eqref{eq:witnessTerm} (unless $|\langle\hat{a}^2\rangle|=p(1-p)(r_{21}+r_{12})^2$, in which case $\mathcal{W}=0$). This is because $\langle\hat{a}^2\rangle$ is too large to cancel in Eq.~\eqref{eq:ORTMixed} and~\eqref{eq:witnessTerm}. In this case the alternate measure $\mathcal{Q}$ (Eq.~\eqref{eq:measureTerm}) and the witness $\mathcal{W}$ always have a common optimal decomposition, and $\mathcal{N}=\mathcal{Q}+\mathcal{W}$. This optimal decomposition of $\hat{\rho}$ is into states $\sqrt{p}\ket{\psi_1}\pm e^{-i\chi}\sqrt{1-p}\ket{\psi_2}$ with probabilities $\frac{1}{2}$. Here, $2\chi\mod2\pi=\arg(\bra{\psi_1}\hat{a}\ket{\psi_2})-\arg(\bra{\psi_2}\hat{a}\ket{\psi_1})$.

Eq.~\eqref{eq:underSqueezed} corresponds to a vanishing nonclassicality witness, $\mathcal{W}=0$. In this case $\mathcal{Q}$ and $\mathcal{W}$ do not have a common optimal decomposition in general, so $\mathcal{N}\neq\mathcal{Q}+\mathcal{W}$ in general (though exceptions may occur, as in mixtures of two neighboring Fock states). $\mathcal{N}$ and $\mathcal{W}$ have a common optimal decomposition, however. This decomposition is into states $\sqrt{p}\ket{\psi_1}+e^{i(\theta-\chi)}\sqrt{1-p}\ket{\psi_2}$, with probabilities $q_\theta$ chosen to nullify the witness term and realize the original density matrix $\hat{\rho}$. For context, it is sufficient to choose symmetric $q_\theta=q_{-\theta}$ satisfying the equations:
\begin{subequations}
\begin{align}
\sum_\theta q_\theta\cos2\theta&=\frac{|\langle\hat{a}^2\rangle|-2r_{21}r_{12}p(1-p)}{p(1-p)(r_{21}^2+r_{12}^2)}\label{eq:cos2theta}\equiv\bar{x}\\
\sum_\theta q_\theta\cos\theta&=0
\end{align}
\end{subequations}
Since $\bar{x}$ in Eq.~\eqref{eq:cos2theta} lies in the range $(-1,1)$, the above equalities are always achievable using weights: $q_0=q_\pi=(\bar{x}+1)/4$ and $q_{\pi/2}=q_{-\pi/2}=(1-\bar{x})/4$.

For comparison, we also provide the analytical expression of the metrological power of these states. 
Applying Eq.~\eqref{eq:MetroPower}--\eqref{eq:fisherEigendecomposition} to rank two states that satisfy Assumptions 1 and 2, one finds: 

\begin{widetext}
\begin{DispWithArrows}<\mathcal{M}(\hat{\rho})=>[format = ll,subequations]
\text{max}[\braket{\hat{n}}+|\braket{\hat{a}^2}|-2p(1-p)(r_{21}+r_{12})^2,0], & \qquad  |\braket{\hat{a}^2}|\geq 4p(1-p)r_{21}r_{12}, \label{eq:M1}\\
\text{max}[\braket{\hat{n}}-|\braket{\hat{a}^2}|-2p(1-p)(r_{12}-r_{21})^2,0] , &\qquad  |\braket{\hat{a}^2}|< 4p(1-p)r_{21}r_{12}.\label{eq:M2}
\end{DispWithArrows}
\end{widetext}
Like $\mathcal{N}$, $\mathcal{M}$ behaves as the piecewise concatenation of two functions in $p$ (assuming a fixed eigenbasis $\{\ket{\psi_1},\ket{\psi_2}\}$). The particular curve $\mathcal{M}$ follows depends on whether the line $p|\langle\hat{a}^2\rangle_1|+(1-p)|\langle\hat{a}^2\rangle_2|$ is above the quadratic $4p(1-p)r_{21}r_{12}$ (see Fig.~\ref{fig:5cases}). 

For a fixed eigenbasis $\{\ket{\psi_1},\ket{\psi_2}\}$, the inequalities governing $\mathcal{N}(\hat{\rho})$ and $\mathcal{M}(\hat{\rho})$ (see Eq. \eqref{eq:overSqueezed}--\eqref{eq:underSqueezed} and Eq. \eqref{eq:M1}--\eqref{eq:M2}) can lead to any of three general scenarios, as depicted in Fig. \ref{fig:5cases}. The solid (blue) and dashed (red) parabolas represent fixed $p(1-p)(r_{21}+r_{12})^2$ and $4p(1-p)r_{21}r_{12}$, respectively, the solid-blue necessarily being above (or coinciding with) the dashed-red because $(r_{21}+r_{12})^2\geq 4r_{21}r_{12}$. If the line $|\langle\hat{a}^2\rangle|=p|\langle\hat{a}^2\rangle|_1+(1-p)|\langle\hat{a}^2\rangle|_2$ is completely above both parabolas, as in dotted (gray) line 1, there are no intervals where Eq. \eqref{eq:underSqueezed} or Eq. \eqref{eq:M2}
hold. If the $|\langle\hat{a}^2\rangle|$ line is partially below the solid (blue) but not the dashed (red) parabola, as in dotted (gray) line 2, there is a single interval $(p_L,p_R)$ where Eq. \eqref{eq:underSqueezed} holds, but no interval where Eq. \eqref{eq:M2} holds. Lastly, if the $|\langle\hat{a}^2\rangle|$ line is partially below both parabolas, there is an interval $(p_L,p_R)$ where Eq. \eqref{eq:underSqueezed} holds, and nested within that an interval $(p_L',p_R')$ where Eq. \eqref{eq:M2} holds. At any $p$ where $|\langle\hat{a}^2\rangle|$ is above the solid (blue) parabola, Eq. \eqref{eq:overSqueezed} and Eq. \eqref{eq:M1} hold, hence $\mathcal{M}=\mathcal{N}$ (all of the nonclassicality translates to metrological advantage). Where $|\langle\hat{a}^2\rangle|$ is below the solid (blue) parabola, $\mathcal{M}\leq\mathcal{N}$, being equal only in specific cases (of which section \ref{sec:indefiniteParity} is one example).

For each of the three scenarios in Fig. \ref{fig:5cases}, we provide at least once instance below: scenario 1 in \ref{sec:squeezed}, scenario 2 in \ref{sec:twoCat} and \ref{sec:squeezed} (same type of state but different parameter regime), and scenario 3 in \ref{sec:twoFock}, \ref{sec:levelSkipping}, and \ref{sec:indefiniteParity}. We find that, even within each general scenario, there are nontrivial variations, demonstrating the variety of ways $\mathcal{N}(\hat{\rho})$ and $\mathcal{M}(\hat{\rho})$ may depend on eigenvalue $p$.
 
\subsection{Mixture of Two Neighboring Fock States}\label{sec:twoFock}
Our general formula recovers the ORT measure for mixtures of two neighboring Fock states $\hat{\rho}_{2F}=p\ket{n+1}\bra{n+1}+(1-p)\ket{n}\bra{n}$. These fall into our special class of rank-2 mixtures, since $\bra{n}\hat{a}\ket{n}=0$ for all Fock states $\ket{n}$ and $\Tr(\hat{\rho}_{2F}\hat{a}^2)=0$. The ORT measure for these states is known from previous works \cite{ge2020evaluating,rogers2024quantifying} to be:
\begin{equation}
\mathcal{N}(\hat{\rho}_{2F})=\underbrace{n+p}_{\langle{n}\rangle}-p(1-p)(n+1).
\label{eq:mixedFock}
\end{equation}
To see how our general formula recovers this, let:
$\ket{\psi_1}=\ket{n+1}$ and $\ket{\psi_2}=\ket{n}$. Then $r_{12}=\bra{n+1}\hat{a}\ket{n}=0$, $r_{21}=\bra{n}\hat{a}\ket{n+1}=\sqrt{n+1}$, and $\langle\hat{a}^2\rangle=p\bra{n+1}\hat{a}^2\ket{n+1}+(1-p)\bra{n}\hat{a}^2\ket{n}=0$. Clearly, $|\langle\hat{a}^2\rangle|< p(1-p)(r_{21}+r_{12})^2$ and Eq.~\eqref{eq:mixedFock} follows from substitution into Eq.~\eqref{eq:underSqueezed}.

For context, the metrological power is:
\begin{equation}
\mathcal{M}(\hat{\rho}_{2F})=\max\left[n+p-2p(1-p)(n+1),0\right],
\end{equation}
and fails to saturate the ORT bound always. $\mathcal{M}<\mathcal{N}$ for all $p$, which is only possible because $|\langle\hat{a}^2\rangle|=0$ for all $p$ (a special case of scenario 3 of Fig. \ref{fig:5cases}).\footnote{We do not consider $p=0$ and $p=1$ to be part of this discussion, since such states would be of lower rank.} Notably, the $n=0$ case corresponds exactly to a single-photon state with loss probability $(1-p)$. $\mathcal{M}$ reaches zero for such mixtures with $\langle\hat{n}\rangle\leq\frac{1}{2}$. This is an example where nonclassicality does not translate to metrological advantage over classical states \cite{ge2020operational}.

\subsection{Mixture of Two Cat States}\label{sec:twoCat}
Cat states can be useful for quantum-enhanced sensing applications, such as phase estimation \cite{stammer2024metrological,gilchrist2004schrodinger,lee2015quantum,zheng2025quantum}. In preparing optical coherent cat states, photon losses lead to mixtures of superposition states of the coherent states~\cite{Hach:94,Glancy:08,Zhang:21,stammer2024metrological}. Here we consider such a mixture of even and odd cat states:
\begin{equation}
\hat{\rho}=p\ket{\text{cat}_+}\bra{\text{cat}_+}+(1-p)\ket{\text{cat}_-}\bra{\text{cat}_-}
\end{equation}
where $\ket{\text{cat}_\pm}=(2\pm 2e^{-2|\alpha|^2})^{-1/2}(\ket{\alpha}\pm\ket{-\alpha})$ are the normalized even and odd cat states, respectively. The cat states are orthogonal, and ``swapped" via the annihilation operator: $\ket{\text{cat}_\mp}\propto\hat{a}\ket{\text{cat}_\pm}$, which together imply $\bra{\text{cat}_\pm}\hat{a}\ket{\text{cat}_\pm}=0$ (Assumption 1 is satisfied). One finds that:
\begin{subequations}
\begin{align}
\bra{\text{cat}_+}\hat{a}\ket{\text{cat}_-}&=\alpha\sqrt{\coth|\alpha|^2}\label{eq:catA}\\
\bra{\text{cat}_-}\hat{a}\ket{\text{cat}_+}&=\alpha\sqrt{\tanh|\alpha|^2}\\
\bra{\text{cat}_+}\hat{a}^2\ket{\text{cat}_+}&=\alpha^2\\
\bra{\text{cat}_-}\hat{a}^2\ket{\text{cat}_-}&=\alpha^2\\
\bra{\text{cat}_+}\hat{n}\ket{\text{cat}_+}&=|\alpha|^2\tanh|\alpha|^2\\
\bra{\text{cat}_-}\hat{n}\ket{\text{cat}_-}&=|\alpha|^2\coth|\alpha|^2.\label{eq:catF}
\end{align}
\end{subequations}
From this, we see that $\bra{\text{cat}_+}\hat{a}^2\ket{\text{cat}_+}$and $\bra{\text{cat}_-}\hat{a}^2\ket{\text{cat}_-}$ have the proper phase to satisfy Assumption 2.

\begin{figure}
\begin{subfigure}{\columnwidth}
  \centering
  \includegraphics[width=\columnwidth]{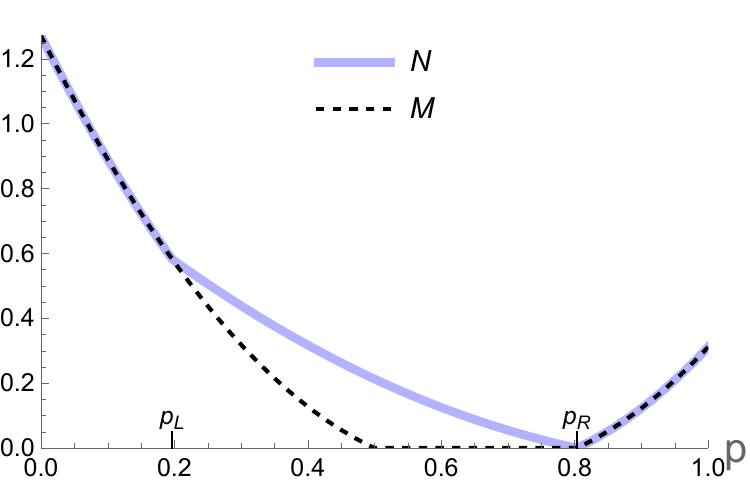}
  \caption{$\alpha=0.5$.}
  \label{fig:lowAlpha}
\end{subfigure}
\hfill
\begin{subfigure}{\columnwidth}
  \centering
  \includegraphics[width=\columnwidth]{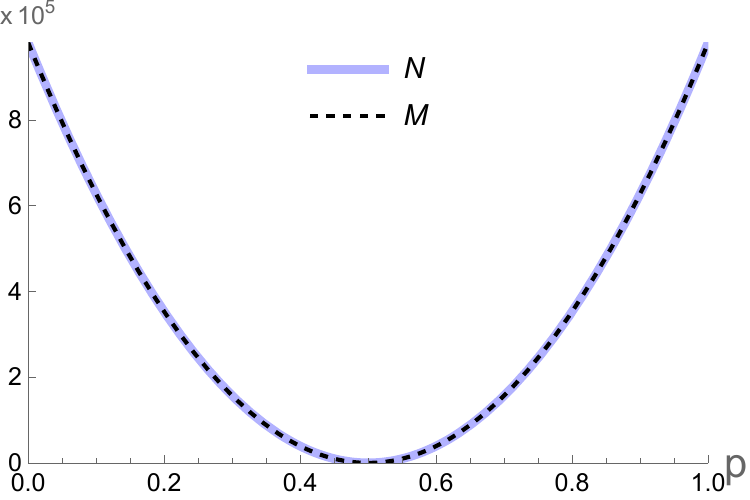}
  \caption{$\alpha=700$.}
  \label{fig:highAlpha}
\end{subfigure}
 \captionsetup{justification=raggedright, singlelinecheck=false}
\caption{The ORT measure $\mathcal{N}$ (solid blue) and the metrological power $\mathcal{M}$ (dashed black) of mixed even and odd cat states, as a function of the population $p$ of the even cat state for (a) $\alpha=0.5$ and for (b) $\alpha=700$. At $p=0$ (odd cat) and $p=1$ (even cat), the nonclassicalities are $\mathcal{N}_{-}=|\alpha|^2(\coth{|\alpha|^2}+1)$ and $\mathcal{N}_{+}=|\alpha|^2(\tanh{|\alpha|^2}+1)$, respectively. $\mathcal{N}_{-}$ and $\mathcal{N}_{+}$ approach one another as $|\alpha|\rightarrow\infty$.}
\label{fig:cats}
\end{figure}

In Fig. \ref{fig:lowAlpha} and \ref{fig:highAlpha} we show the ORT measure and metrological power (as a function of $p$) for small and large $\alpha$. In the region between $p_L=(1-e^{-2|\alpha|^2})/2$ and $p_R=(1+e^{-2|\alpha|^2})/2$, $\mathcal{N}(\hat{\rho})$ obeys Eq.~\eqref{eq:underSqueezed} and $\mathcal{M}<\mathcal{N}$; outside this region, $\mathcal{N}$ obeys Eq.~\eqref{eq:overSqueezed} and $\mathcal{M}=\mathcal{N}$. Since the line $p|\langle\hat{a}^2\rangle_1|+(1-p)|\langle\hat{a}^2\rangle_2|$ is always above the curve $4p(1-p)r_{21}r_{12}$ (scenario 2 in Fig. \ref{fig:5cases}), $\mathcal{M}$ is always given by Eq. \eqref{eq:M1}. Interestingly, the nonclassicality vanishes completely at $p_R$ (the state becoming equivalent to a mixture of the coherent states $\ket{\alpha}$ and $\ket{-\alpha}$), and the metrological power vanishes in the interval $(\frac{1}{2},p_R)$. For large $\alpha$, the middle $(p_L,p_R)$ becomes only a small neighborhood of $p=\frac{1}{2}$, and the ORT measure and metrological power behave essentially as a symmetric quadratic about $p=\frac{1}{2}$. We note that this vanishing metrological power at $p=\frac{1}{2}$ indicates no quantum sensing advantage, as the $p=\frac{1}{2}$ state is the natural limit of a pure cat with appreciable loss \cite{stammer2024metrological}.

\subsection{Mixture of Squeezed Vacuum and Photon-subtracted Squeezed Vacuum States}\label{sec:squeezed}
Next we consider mixtures of squeezed vacuum and photon-subtracted squeezed vacuum, represented by:
\begin{equation}
\hat{\rho}=p\ket{\text{SV}}\bra{\text{SV}}+(1-p)\ket{\text{SV}_-}\bra{\text{SV}_-},
\end{equation}
where $\ket{\text{SV}}=\hat{S}\ket{0}$ with $\hat{S}=\exp[\frac{1}{2}(\gamma e^{-2i\mu}\hat{a}^2-\gamma e^{2i\mu}\hat{a}^{\dagger 2})]$ is a squeezed vacuum state and $\ket{\text{SV}_-}=(-ie^{-i\mu}\csch\gamma)\hat{a}\ket{\text{SV}}$ is a photon-subtracted squeezed vacuum state. $\gamma>0$ determines the squeezing strength and $\mu$ determines the squeezed quadrature. Squeezed vacuum is known for its theoretical optimality for phase-sensing and demonstrated usefulness in gravitational-wave interferometry \cite{caves1981quantum,Lang13,ligo2011gravitational,aasi2013enhanced}, while the nonclassical properties of photon-subtracted squeezed states were themselves investigated in Ref. \cite{wang2012photon,thapliyal2017comparison}. Squeezed vacuum is a superposition of only even Fock states $\{\ket{2n}\}$, while photon-subtracted squeezed vacuum is a superposition of only odd Fock states $\{\ket{2n+1}\}$, so $\braket{\text{SV}|\text{SV}_-}=0$, $\bra{\text{SV}}\hat{a}\ket{\text{SV}}=0$, and $\bra{\text{SV}_-}\hat{a}\ket{\text{SV}_-}=0$ (Assumption 1 is satisfied).

One finds that:
\begin{subequations}
\begin{align}
\bra{\text{SV}}\hat{a}\ket{\text{SV}_-}&=ie^{i\mu}\cosh\gamma\label{eq:squeezedSetA}\\
\bra{\text{SV}_-}\hat{a}\ket{\text{SV}}&=ie^{i\mu}\sinh\gamma\\
\bra{\text{SV}}\hat{a}^2\ket{\text{SV}}&=-e^{2i\mu}\sinh\gamma\cosh\gamma\\
\bra{\text{SV}_-}\hat{a}^2\ket{\text{SV}_-}&=-3e^{2i\mu}\sinh\gamma\cosh\gamma\label{eq:squeezedSetD}\\
\bra{\text{SV}}\hat{n}\ket{\text{SV}}&=\sinh^2\gamma\\
\bra{\text{SV}_-}\hat{n}\ket{\text{SV}_-}&=2\sinh^2\gamma+\cosh^2\gamma.
\end{align}
\end{subequations}
From Eq.~\eqref{eq:squeezedSetA}--\eqref{eq:squeezedSetD} we see that Assumption 2 is satisfied, since $-e^{2i\mu}=(ie^{i\mu})^2$.

\begin{figure}
  \centering
  \includegraphics[width=\columnwidth]{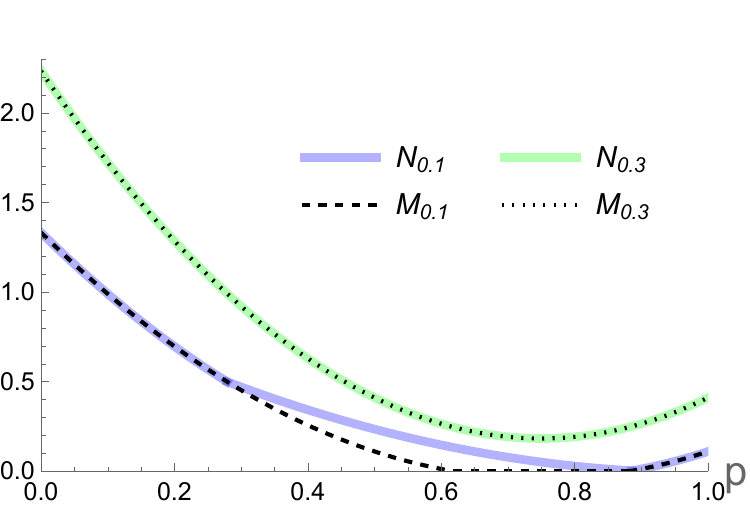}
  \label{fig:lowGamma}
 \captionsetup{justification=raggedright, singlelinecheck=false}
\caption{ORT measure $\mathcal{N}$ and metrological power $\mathcal{M}$ of mixtures of squeezed vacuum and photon-subtracted squeezed vacuum for a squeezing parameter below ($\gamma=0.1$) and above ($\gamma=0.3$) $\gamma_{\text{crit}}\approx0.19193$, as a function of the population $p$ of the squeezed vacuum state. To clarify, the ORT measure for states with $\gamma=0.1$ does not reach zero for any $p$.}
\label{fig:squeezedMixtureLowGamma}
\end{figure}

One can show that Eq.~\eqref{eq:overSqueezed} always gives the nonclassicality if the squeezing parameter $\gamma$ is above $\gamma_{\text{crit}}=[\ln(1+2/\sqrt{3})]/4\approx0.19193$. In this case, $\mathcal{M}=\mathcal{N}$, regardless of $p$. This is an example of scenario 1 in Fig. \ref{fig:5cases}. For $\gamma<\gamma_\text{crit}$, there will be an interval $(p_L,p_R)$ (dependent on $\gamma$) where $\mathcal{M}<\mathcal{N}$, however $\mathcal{M}=\mathcal{N}$ outside that interval. This is an example of scenario 2 in Fig. \ref{fig:5cases}. Both $\gamma<\gamma_\text{crit}$ and $\gamma>\gamma_\text{crit}$ cases are plotted in Fig. \ref{fig:squeezedMixtureLowGamma}.

\subsection{Mixture of Level-Skipping States}\label{sec:levelSkipping}

As an example of scenario 3 in Fig. \ref{fig:5cases}, we next introduce mixtures of ``level-skipping states." By such mixtures, we mean:
\begin{equation}
\hat{\rho}=p\ket{\phi_1}\bra{\phi_1}+(1-p)\ket{\phi_2}\bra{\phi_2},
\end{equation}
where
$\ket{\phi_2}=\alpha_2\ket{n}+\beta_2\ket{n+2}$ is normalized and, without loss of generality, $\beta_2$ is complex while $\alpha_2$ is positive and real. We take the other level-skipping state, $\ket{\phi_1}$, to be related to $\ket{\phi_2}$ by $\ket{\phi_1}=(n+2|\beta_2|^2)^{-1/2}\hat{a}\ket{\phi_2}$. This type of mixture is relevant to experimental generation of a superposition of odd photon number states under weak photon losses ~\cite{Dakna:97,Nielsen:06}.

Since each $\ket{\phi_i}$ exclusively involves even or odd Fock states, $\bra{\phi_i}\hat{a}\ket{\phi_i}=0$ -- Assumption 1 is satisfied. One also finds that:
\begin{subequations}
\begin{align}
    \bra{\phi_1}\hat{a}\ket{\phi_2}&=\sqrt{n+2|\beta_2|^2} \label{eq:skippedSetA}\\
    \bra{\phi_2}\hat{a}\ket{\phi_1}&=\frac{\alpha_2\beta_2\sqrt{(n+2)(n+1)}}{\sqrt{n+2|\beta_2|^2}}\\
    \bra{\phi_1}\hat{a}^2\ket{\phi_1}
    &=\frac{\alpha_2\beta_2n\sqrt{(n+2)(n+1)}}{n+2|\beta_2|^2}\\
    \bra{\phi_2}\hat{a}^2\ket{\phi_2}
    &=\alpha_2\beta_2\sqrt{(n+2)(n+1)} \label{eq:skippedSetD}\\
    \bra{\phi_1}\hat{n}\ket{\phi_1}&=n-\frac{n}{n+2|\beta_2|^2}\alpha_2^2+\frac{n+2}{n+2|\beta_2|^2}|\beta_2|^2\\
    \bra{\phi_2}\hat{n}\ket{\phi_2}&=n+2|\beta_2|^2.
\end{align}
\end{subequations}
Analysis of the phases in Eq.~\eqref{eq:skippedSetA}--\eqref{eq:skippedSetD} shows that Assumption 2 is also satisfied. 

\begin{figure}
\begin{subfigure}{\columnwidth}
  \centering
  \includegraphics[width=\textwidth]{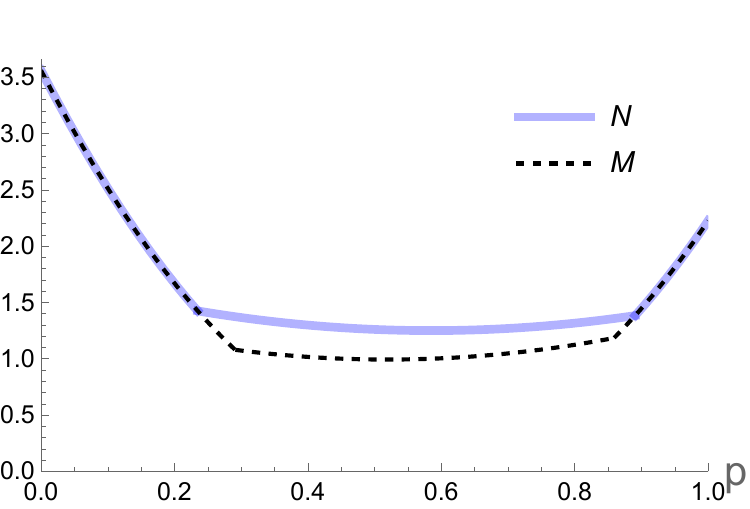}
  \caption{$n=1,|\beta_2|=\sqrt{0.75}$}
  \label{fig:rank2ORT}
\end{subfigure}
\hfill
\begin{subfigure}{\columnwidth}
  \centering
  \includegraphics[width=\textwidth]{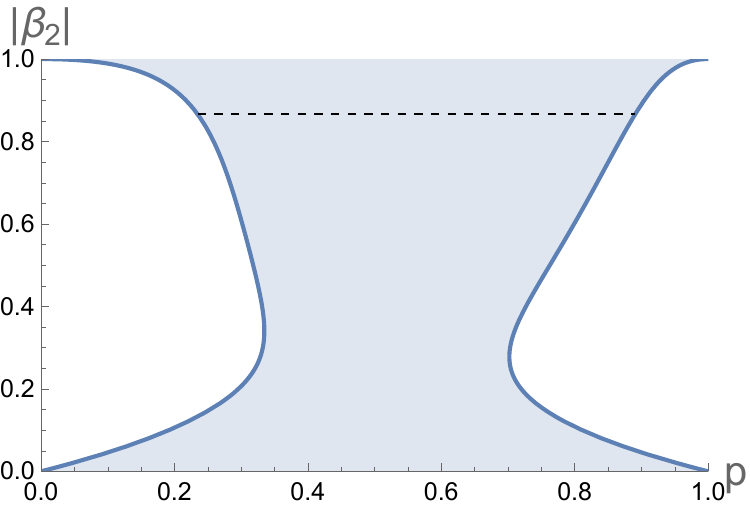}
  \caption{}
  \label{fig:rank2ORT_bvar}
\end{subfigure}
 \captionsetup{justification=raggedright, singlelinecheck=false}
\caption{Fig. \ref{fig:rank2ORT} shows the ORT measure $\mathcal{N}$ (solid, blue) and metrological power $\mathcal{M}$ (dashed, black) of mixtures of level-skipped states with $n=1$ and $|\beta_2|=\sqrt{0.75}$, as a function of the population $p$ of $\ket{\phi_1}$. Fig. \ref{fig:rank2ORT_bvar} shows the variation of $(p_L,p_R)$ with $|\beta_2|$ when $n=1$. The black dotted line represents $(p_L,p_R)\approx(0.234823,0.890667)$ for the case shown in \ref{fig:rank2ORT}.}
\label{fig:fullrank2ORT}
\end{figure}

Fig. \ref{fig:rank2ORT} shows the ORT measure and metrological power for such $\hat{\rho}$ as a function of $p$ with $n=1$ and $|\beta_2|=\sqrt{0.75}$ ($\mathcal{N}$ and $\mathcal{M}$ are independent of arg($\beta_2$)). For an interval $(p_L,p_R)$, the ORT measure is given by Eq. (\ref{eq:underSqueezed}) and $\mathcal{M}<\mathcal{N}$; outside this interval, the ORT measure is given by Eq. (\ref{eq:overSqueezed}) and $\mathcal{M}=\mathcal{N}$. The precise interval $(p_L,p_R)$ depends on $|\beta_2|$ and $n$. Fig. \ref{fig:rank2ORT_bvar} shows its variation with $|\beta_2|$. It can be shown that in the limit $n \to \infty, |\beta_2|=\frac{1}{\sqrt{2}}$ gives the smallest interval: $(p_L,p_R)=(\frac{1}{3},\frac{2}{3})$. Moreover, the length of this interval is never zero: the ORT measure for these mixtures will always access Eq. (\ref{eq:underSqueezed}) for some values of $p$. 

Note that the metrological power has its own piecewise behavior, characterized by an interval $(p_L',p_R')$ nested within $(p_L,p_R)$, as exhibited in Fig. \ref{fig:rank2ORT}. This is a symptom of scenario 3 in Fig. \ref{fig:5cases}.



\subsection{Mixture of Indefinite Parity States}\label{sec:indefiniteParity}
All mixtures considered so far have eigenstates with definite and opposite parity (that is, they are eigenstates of $\hat{P}=\sum_n(-1)^n\ket{n}\bra{n}$ with different eigenvalues). We now consider a mixture that is not of this 
type.
\begin{equation}
\hat{\rho}=p\ket{\Xi_1}\bra{\Xi_1}+(1-p)\ket{\Xi_2}\bra{\Xi_2} 
\end{equation}
where \begin{align*}
  \ket{\Xi_1}&=B(\sqrt{n+1}\ket{n+2}+ze^{ib}\ket{n+1}-\sqrt{n+2}e^{i2b}\ket{n})\\
  \ket{\Xi_2}&=A(\sqrt{n+1}\ket{n+2}+ye^{ia}\ket{n+1}-\sqrt{n+2}e^{i2a}\ket{n}).
\end{align*}
$A,B$ are complex numbers such that both states are normalized, and $y,z,a,b$ are real numbers with $y,z>0$. This selection of coefficients implies $\bra{\Xi_i}\hat{a}\ket{\Xi_i}=0$. We also require $\braket{\Xi_1|\Xi_2}=0$. One way to achieve this is to impose that $a-b=k\pi$ where $k$ is odd, and $yz=2n+3$. With these, one finds that
\begin{subequations}
    \begin{align}
    \bra{\Xi_1}\hat{a}\ket{\Xi_2}&=\sqrt{(n+1)(n+2)}AB^*(z+y)e^{-ib}\\
    \bra{\Xi_2}\hat{a}\ket{\Xi_1}&=\sqrt{(n+1)(n+2)}BA^*(z+y)e^{-ia}\\
    \bra{\Xi_1}\hat{a}^2\ket{\Xi_1}&=|B|^2(n+1)(n+2)e^{-i(2b+k\pi)}\\
    \bra{\Xi_2}\hat{a}^2\ket{\Xi_2}&=|A|^2(n+1)(n+2)e^{-i(2a-k\pi)}\\
    \bra{\Xi_1}\hat{n}\ket{\Xi_1}&=|B|^2(2n^2+5n+2+(n+1)z^2)\\
    \bra{\Xi_2}\hat{n}\ket{\Xi_2}&=|A|^2(2n^2+5n+2+(n+1)y^2).
\end{align}
\end{subequations}
These expressions and $a-b=k\pi$ imply that such mixtures satisfy Assumption 2. 

\begin{figure}
  \centering
   \captionsetup{justification=raggedright, singlelinecheck=false}
  \includegraphics[width=\columnwidth]{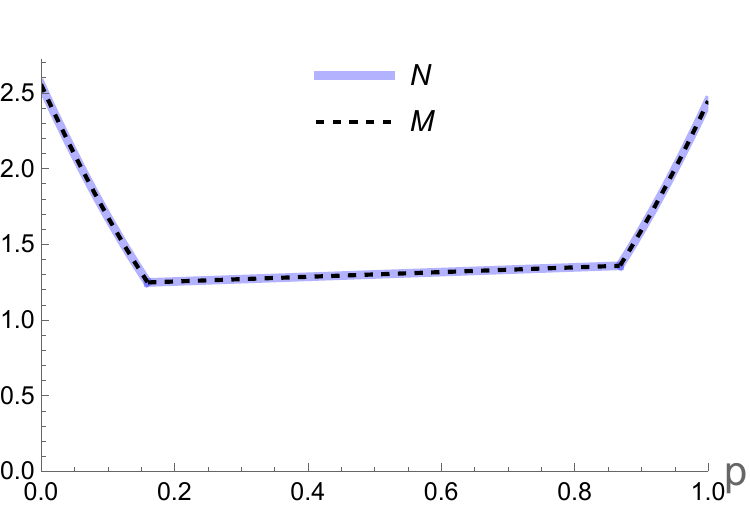}
  \caption{ORT measure $\mathcal{N}$ (solid, blue) and metrological power $\mathcal{M}$ (dashed, black) of mixtures of certain indefinite parity states with $n=1$ and $y=2$, as a function of the population $p$ of $\ket{\Xi_1}$. Here $(p_L,p_R)\approx(0.160063,0.867715).$}
  \label{fig:ORT_indef_parity}
\end{figure}

Fig. \ref{fig:ORT_indef_parity} shows the ORT measure and metrological power for such $\hat{\rho}$ as a function of $p$ with $n=1$ and $y=2$. For an interval $(p_L,p_R)$, the ORT measure is given by Eq. (\ref{eq:underSqueezed}); outside this interval, the ORT measure is given by Eq. (\ref{eq:overSqueezed}). As before, the precise interval $(p_L,p_R)$ depends on $n$ and $y$. Regardless of these choices, such mixtures always access both Eq. (\ref{eq:overSqueezed}) and Eq. (\ref{eq:underSqueezed}) as $p$ varies. Moreover, for these states $r_{12}=r_{21}$, which necessitates that the metrological power always saturate the ORT bound ($\mathcal{M}=\mathcal{N}$) and that both are linear in $p$ in the interval $(p_L,p_R)$. This example follows scenario 3 in Fig. \ref{fig:5cases} (note that when $r_{12}=r_{21}$, the red and blue curves in Fig. \ref{fig:5cases} coincide). It thus provides a contrast to mixtures of neighboring Fock states (\ref{sec:twoFock}) and level-skipping states (\ref{sec:levelSkipping}), which are also instances of scenario 3, but have $\mathcal{M}<\mathcal{N}$ in an interval.

\section{Adding Nonzero Coherences}\label{sec:addingCoherence}
Now we allow for partial coherence between the former eigenstates $\ket{\psi_1}$ and $\ket{\psi_2}$.
\begin{equation}
\begin{split}
\hat{\rho}&=p\ket{\psi_1}\bra{\psi_1}+fe^{i\chi}\sqrt{p(1-p)}\ket{\psi_1}\bra{\psi_2}\\
&+fe^{-i\chi}\sqrt{p(1-p)}\ket{\psi_2}\bra{\psi_1}+(1-p)\ket{\psi_2}\bra{\psi_2}.
\end{split}
\label{eq:partiallyCoherentMixture}
\end{equation}
Here, $f$ satisfying $0\leq f\leq1$ measures the \textit{coherence strength}  between the states $\ket{\psi_1}$ and $\ket{\psi_2}$. $f=1$ corresponds to a pure superposition $\sqrt{p}\ket{\psi_1}+e^{-i\chi}\sqrt{1-p}\ket{\psi_2}$, whereas $f=0$ corresponds to a completely incoherent mixture of $\ket{\psi_1}$ and $\ket{\psi_2}$. $|f|>1$ is not allowed, since this would produce an unphysical negative eigenvalue for $\hat{\rho}$. Since $\ket{\psi_1}$ and $\ket{\psi_2}$ are no longer the eigenstates of $\hat{\rho}$, we will refer to them as the \textit{special basis states.}


The following assumptions on $\hat{\rho}$ allow the ORT measure to take a convenient form.

\textbf{Assumption 1*: The special basis states are centered at the origin in phase space.} By this we mean that $\bra{\psi_1}\hat{a}\ket{\psi_1}=\bra{\psi_2}\hat{a}\ket{\psi_2}=0$.

\textbf{Assumption 2*: The special basis states satisfy $\arg(\bra{\psi_1}\hat{a}^2\ket{\psi_1})=\arg(\bra{\psi_2}\hat{a}^2\ket{\psi_2})=[\arg(\bra{\psi_1}\hat{a}\ket{\psi_2})+\arg(\bra{\psi_2}\hat{a}\ket{\psi_1})] \mod 2\pi$.} Again, we consider any instance of $\arg(0)$ to be its own free variable, and we consider the equations satisfied if some choice yields equality (the same goes for Assumption 3 to follow).

\textbf{Assumption 3: The phase of the coherence satisfies} $2\chi\mod2\pi=\arg(\bra{\psi_1}\hat{a}\ket{\psi_2})-\arg(\bra{\psi_2}\hat{a}\ket{\psi_1})$. Note that one may always perform a change of basis: $\ket{\psi_1}\rightarrow e^{i\theta_1}\ket{\psi_1}$ and $\ket{\psi_2}\rightarrow e^{i\theta_2}\ket{\psi_2}$, such that $\arg(\bra{\psi_1}\hat{a}\ket{\psi_2})=\arg(\bra{\psi_2}\hat{a}\ket{\psi_1})$ (this does not alter the validity of Assumptions 1* and 2*). With this choice of basis, Assumption 3 is the condition that $\chi=0$ or $\pi$.

\textbf{Assumption 4: The special basis states and $\chi$ satisfy $e^{-i\chi}\bra{\psi_1}\hat{a}^2\ket{\psi_2}+e^{i\chi}\bra{\psi_2}\hat{a}^2\ket{\psi_1}=0$.} This makes $\langle\hat{a}^2\rangle$ invariant under $f$.\footnote{One could make the weaker assumption that $\arg(\langle\hat{a}^2\rangle)$ is invariant under $f$, but that is beyond the scope of this paper.}

Assumptions 3 and 4 are essentially phase-matching conditions that simplify the calculation of $\mathcal{N}$, enabling us to quantify the effect of decoherence (particularly dephasing) on nonclassicality under certain conditions. Generally, we consider cases where the special basis states $\ket{\psi_1}$ and $\ket{\psi_2}$ have definite and opposite parity $\hat{P}=\sum_n(-1)^n\ket{n}\bra{n}$. In this case, Assumption 1* is always satisfied, as is Assumption 4. In addition, $\langle\hat{n}\rangle=p\langle\hat{n}\rangle_1+(1-p)\langle\hat{n}\rangle_2$ and $|\langle\hat{a}^2\rangle|=p|\langle\hat{a}^2\rangle_1|+(1-p)|\langle\hat{a}^2\rangle_2|$, as before (these quantities are independent of $f$). Lastly, bosonic dephasing can explain decreases in the magnitude of $f$ via the operation: $\mathcal{D}(\hat{\rho})=\frac{1}{2}\left[\left(1-f'/f\right)e^{-i\hat{n}\pi}\hat{\rho}e^{i\hat{n}\pi}+\left(1+f'/f\right)\hat{\rho}\right]$. Thus, decreasing $f$ can never increase nonclassicality, by Eq. \eqref{eq:generalBDCBound}.

Making Assumptions 1*, 2*, 3, and 4, the ORT measure obeys:
\begin{widetext}
\begin{DispWithArrows}<\mathcal{N}(\hat{\rho})=>[format = ll,subequations]
\langle\hat{n}\rangle+|\langle\hat{a}^2\rangle|-2p(1-p)(r_{21}+r_{12})^2 & \quad  |\langle\hat{a}^2\rangle|\geq p(1-p)(r_{21}+r_{12})^2 \label{eq:overSqueezed2} \\
\langle\hat{n}\rangle+\frac{-1}{r_{21}^2+r_{12}^2}\left(2r_{21}r_{12}|\langle\hat{a}^2\rangle|+(r_{21}^2-r_{12}^2)^2p(1-p)\right) &\quad  |\langle\hat{a}^2\rangle|< p(1-p)(r_{21}+r_{12})^2\cap |\langle\hat{a}^2\rangle|\geq\eta(f,p)\label{eq:underSqueezedA} \\
\langle\hat{n}\rangle-|\langle\hat{a}^2\rangle|+(2f^2-2)p(1-p)(r_{21}-r_{12})^2 &\quad |\langle\hat{a}^2\rangle|< p(1-p)(r_{21}+r_{12})^2\cap |\langle\hat{a}^2\rangle|<\eta(f,p),\label{eq:underSqueezedB}
\end{DispWithArrows}
\end{widetext}
where we have defined:
\begin{equation}
\eta(f,p)=p(1-p)[(2f^2-1)(r_{21}^2+r_{12}^2)+2r_{21}r_{12}].
\label{eq:boundingEta}
\end{equation}
Note that $\eta(1,p)=p(1-p)(r_{21}+r_{12})^2$. See Appendix \ref{sec:appendixDerivation} for details of the derivation.

Like Eq.~\eqref{eq:overSqueezed}, Eq.~\eqref{eq:overSqueezed2} corresponds to a nonzero nonclassicality witness, $\mathcal{W}$ in Eq.~\eqref{eq:witnessTerm}, because $\langle\hat{a}^2\rangle$ is too large to cancel in Eq.~\eqref{eq:ORTMixed} and~\eqref{eq:witnessTerm}. In this case the alternate measure $\mathcal{Q}$ (Eq.~\eqref{eq:measureTerm}) and the witness $\mathcal{W}$ always have a common optimal decomposition, and $\mathcal{N}=\mathcal{Q}+\mathcal{W}$. This optimal decomposition of $\hat{\rho}$ is into states $\sqrt{p}\ket{\psi_1}\pm e^{-i\chi}\sqrt{1-p}\ket{\psi_2}$ with probabilities $(1\pm f)/2$. Interestingly, the value of $f$ does not affect the ORT measure in this regime. One can even take $f=1$, which is the pure state $\sqrt{p}\ket{\psi_1}+e^{-i\chi}\sqrt{1-p}\ket{\psi_2}$.

Like Eq.~\eqref{eq:underSqueezed}, Eq.~\eqref{eq:underSqueezedA} corresponds to a vanishing nonclassicality witness, $\mathcal{W}=0$. In this case $\mathcal{Q}$ and $\mathcal{W}$ do not have a common optimal decomposition in general, so $\mathcal{N}\neq\mathcal{Q}+\mathcal{W}$ in general. $\mathcal{N}$ and $\mathcal{W}$ have a common optimal decomposition, however. This decomposition is into states $\sqrt{p}\ket{\psi_1}+e^{i(\theta-\chi)}\sqrt{1-p}\ket{\psi_2}$, with probabilities $q_\theta$ chosen to nullify the witness term and realize the original density matrix $\hat{\rho}$. For context, it is sufficient to choose symmetric $q_\theta=q_{-\theta}$ satisfying Equations~\eqref{eq:cos2theta} and:
\begin{equation}
\sum_\theta q_\theta\cos\theta=f.
\label{eq:cosThetaAgain}
\end{equation}
$\bar{x}$ in Eq.~\eqref{eq:cos2theta} lies in the range $(2f^2-1,1)$, the lower bound coming from the condition $|\langle\hat{a}^2\rangle|\geq\eta(f,p)$ for Eq.~\eqref{eq:underSqueezedA}. Equations~\eqref{eq:cos2theta} and~\eqref{eq:cosThetaAgain} may always be satisfied by choosing the four angles $\theta_{\pm,\pm}=\pm\arccos({f\pm\sqrt{[\bar{x}-(2f^2-1)]/2}})$ with weights $q_{\pm,\pm}=1/4$ each.

Eq.~\eqref{eq:underSqueezedB} corresponds to a nonzero nonclassicality witness, $\mathcal{W}$ in Eq.~\eqref{eq:witnessTerm}, since $\langle\hat{a}^2\rangle$ is too small to cancel in Eq.~\eqref{eq:ORTMixed} and~\eqref{eq:witnessTerm}. For $\mathcal{N}$ and $\mathcal{W}$, an optimal decomposition of $\hat{\rho}$ is into states $\sqrt{p}\ket{\psi_1}+\sqrt{1-p}e^ {i(\pm\theta_f-\chi)}\ket{\psi_2}$, with probabilities $\frac{1}{2}$ and $\theta_f=\arccos(f)$. $\mathcal{N}$ and $\mathcal{W}$ do not share an optimal decomposition with $\mathcal{Q}$, in general. Lastly, we note that if $|\langle\hat{a}^2\rangle|<p(1-p)(r_{21}+r_{12})^2$, there is always an $f$-interval $(f_L,1)$ where Eq.~\eqref{eq:underSqueezedB} holds, since $\eta(1,p)=p(1-p)(r_{21}+r_{12})^2$ and $\eta(f,p)$ increases monotonically in $f$ (for positive $f$).

One may ask whether we can change the phase of the off-diagonal elements (keeping the basis completely fixed), without affecting the ORT measure. First, we note that Assumption 3 allows two different values of $\chi$: $\chi_0$ and $\chi_0+\pi$. Both yield the same value of the ORT measure (assuming 1*, 2*, and 4 are also met). Second, if the special basis states are Fock states, $\ket{\psi_1}=\ket{n_1}$ and $\ket{\psi_2}=\ket{n_2}$, changing $\chi$ is equivalent to applying a well-defined phase shift, which is a classical operation that does not change the measure (see Appendix \ref{sec:appendixPhaseShift}).

\begin{figure}
    \centering
    \includegraphics[width=0.9\linewidth]{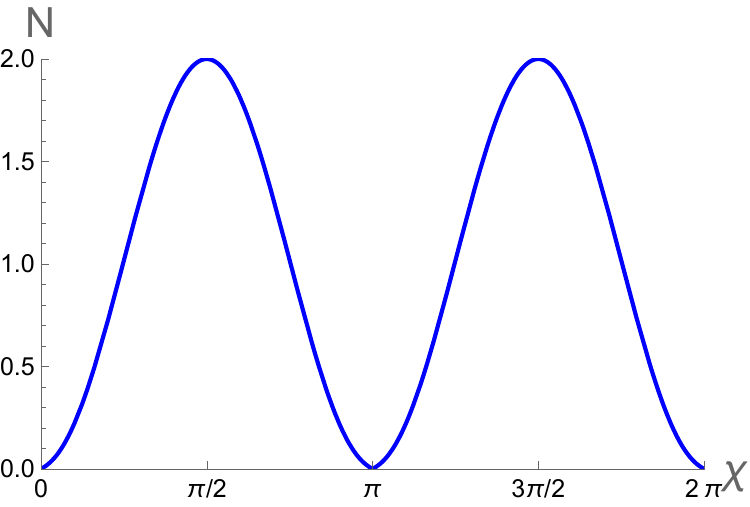}
     \captionsetup{justification=raggedright, singlelinecheck=false}
    \caption{ORT measure $\mathcal{N}$ of a pure superposition of even and odd cat states with $\alpha=1$, as a function of the complex phase $\chi$ (see Eq.~\eqref{eq:pureCatQubit}). The superposition is tuned such that the nonclassicality vanishes at $\chi=0$ and $\chi=\pi$. The maximum $\mathcal{N}$ for this state is $2|\alpha|^2=2$ and occurs at $\frac{\pi}{2}$ and $\frac{3\pi}{2}$. Since $\mathcal{M}=\mathcal{N}$ for pure states, this plot also shows the metrological power.}
    \label{fig:pureCatQubit}
\end{figure}

However, if the special basis states are not Fock states, the ORT measure will generally depend on $\chi$ (hence the importance of Assumption 3). As confirmation of this, one may consider the pure state:
\begin{equation}
\ket{\text{cat}_{qb}}=\sqrt{\frac{1+e^{-2|\alpha|^2}}{2}}\ket{\text{cat}_+}+e^{-i\chi}\sqrt{\frac{1-e^{-2|\alpha|^2}}{2}}\ket{\text{cat}_-}.
\label{eq:pureCatQubit}
\end{equation}
Note that $\ket{\text{cat}_{qb}}$ corresponds to a $f=1$ subcase of Eq.~\eqref{eq:partiallyCoherentMixture}, with $\ket{\text{cat}_\pm}$ as the special basis states $\ket{\psi_1}$ and $\ket{\psi_2}$. Assumptions 1*, 2* and 4 are always satisfied, regardless of $\chi$. Assumption 3, however, is only satisfied by $\chi=0$ or $\pi$. According to Eq.~\eqref{eq:overSqueezed2}--\eqref{eq:underSqueezedB}, these two cases must yield the same nonclassicality, and indeed $\mathcal{N}=0$ for both, since  $\ket{\text{cat}_{qb}}=\ket{\alpha}$ at $\chi=0$ and $\ket{\text{cat}_{qb}}=\ket{-\alpha}$ at $\chi=\pi$. Using the formula for pure states, Eq.~\eqref{eq:pureStateORT}, we can calculate the ORT measure for all $\chi$, and in general $\mathcal{N}\neq0$ and varies with $\chi$, as shown in Fig. \ref{fig:pureCatQubit}.

\subsection{Partially-coherent Mixture of $\ket{n+1}$ and $\ket{n}$}\label{sec:partiallyCoherent2Fock}
As an application, let us consider mixed states of the form:
\begin{equation}
\begin{split}
\hat{\rho}&=p\ket{n+1}\bra{n+1}+fe^{i\chi}\sqrt{p(1-p)}\ket{n+1}\bra{n}\\
&+fe^{-i\chi}\sqrt{p(1-p)}\ket{n}\bra{n+1}+(1-p)\ket{n}\bra{n}.
\end{split}
\label{eq:partiallyCoherent2Fock}
\end{equation}
This state can describe the nonclassical light generated under experimental imperfections~\cite{Loredo:2019aa}. Note that, in this case, any phase $\chi$ will give the same value of the ORT measure, since all values of $\chi$ are related by definite phase shifts that do not affect the ORT measure (see Appendix~\eqref{sec:appendixPhaseShift}). One can thus ignore Assumption 3. Assumptions 1*, 2*, and 4 are also satisfied, and the ORT measure is:
\begin{equation}
\mathcal{N}(\hat{\rho})=p+n+\max(-1,2f^2-2)p(1-p)(n+1).
\label{eq:partiallyCoherentMixture2FockStates}
\end{equation}

\begin{figure}
    \centering
    \includegraphics[width=\columnwidth]{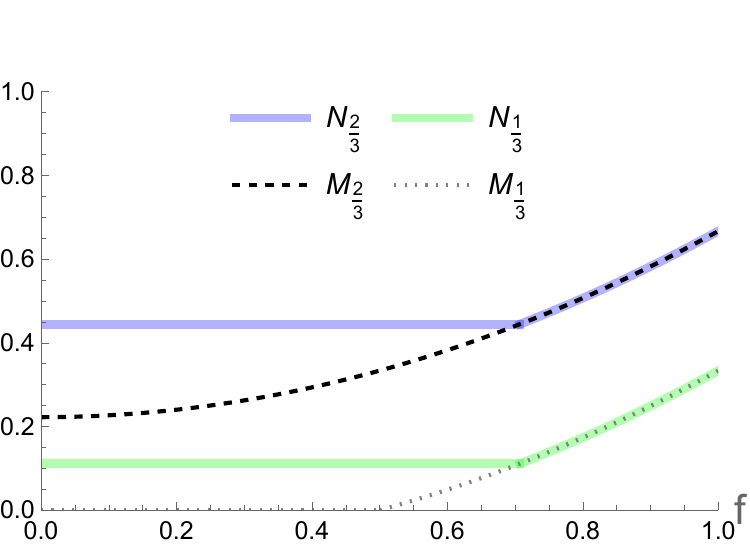}
     \captionsetup{justification=raggedright, singlelinecheck=false}
    \caption{ORT measure $\mathcal{N}$ and metrological power $\mathcal{M}$ for a partially-coherent mixture of nearest-neighbor Fock states $\ket{1}$ and $\ket{0}$, for different choices $p=\frac{2}{3}$ and $p=\frac{1}{3}$ of $\bra{1}\hat{\rho}\ket{1}$, as a function of the coherence strength $f$ (see Eq.~\eqref{eq:partiallyCoherent2Fock}).}
    \label{fig:partiallyCoherentMixture2FockStates}
\end{figure}
The $n=0$ case is plotted in Fig. \ref{fig:partiallyCoherentMixture2FockStates}, as a function of the photon number coherence strength $f$. Interestingly, the ORT measure is constant with respect to $f$ for values of $f<\frac{1}{\sqrt{2}}$ and does not match the metrological power there (this holds regardless of $n$). This regime corresponds to a vanishing nonclassicality witness $\mathcal{W}$ (Eq.~\eqref{eq:underSqueezedA} is valid). For $f>\frac{1}{\sqrt{2}}$, $\mathcal{N}$ is quadratic in $f$, Eq.~\eqref{eq:underSqueezedB} holds, and metrological power equals nonclassicality. The transition across $f=\frac{1}{\sqrt{2}}$ bears some resemblance to entanglement sudden death, whereby entanglement reaches a value of zero in finite time \cite{yu2004finite,yu2009sudden,zyczkowski2001dynamics,Diosi2003,dodd2004disentanglement}. One can imagine a dephasing process $f=e^{-\gamma t}$ such that the pure state $\sqrt{p}\ket{n+1}+e^{-i\chi}\sqrt{1-p}\ket{n}$ evolves to the completely incoherent mixture $p\ket{n+1}\bra{n+1}+(1-p)\ket{n}\bra{n}$ in infinite time, but the nonclassicality reaches its final value in finite time $t=(2\gamma)^{-1}\ln2$ (Fig. \ref{fig:dephasing} depicts this) . However, here the nonclassicality does not completely vanish, only the nonclassicality witness $\mathcal{W}$. Lastly, we find that for $p\leq\frac{1}{2}$, the metrological power vanishes for insufficient coherence $f\leq\sqrt{\frac{1-2p}{2-2p}}$. The metrological power is calculated using diagonalization and Eq.~\eqref{eq:MetroPower} and~\eqref{eq:fisherEigendecomposition}.

\subsection{Partially-Coherent Mixture of Cat States}
Next, we consider a partially-coherent mixture of even and odd cat states:
\begin{equation}
\begin{split}
\hat{\rho}&=p\ket{\text{cat}_+}\bra{\text{cat}_+}+f\sqrt{p(1-p)}\ket{\text{cat}_+}\bra{\text{cat}_-}\\
&+f\sqrt{p(1-p)}\ket{\text{cat}_-}\bra{\text{cat}_+}+(1-p)\ket{\text{cat}_-}\bra{\text{cat}_-}.
\end{split}
\label{eq:partiallyCoherent2Cat}
\end{equation}
Using Eq.~\eqref{eq:catA}--\eqref{eq:catF}, one can show that this mixture obeys Assumptions 1*, 2*, 3, and 4 with special basis states $\ket{\text{cat}_\pm}$.

\begin{figure}
    \centering
     \captionsetup{justification=raggedright, singlelinecheck=false}
     \includegraphics[width=1.05\columnwidth]{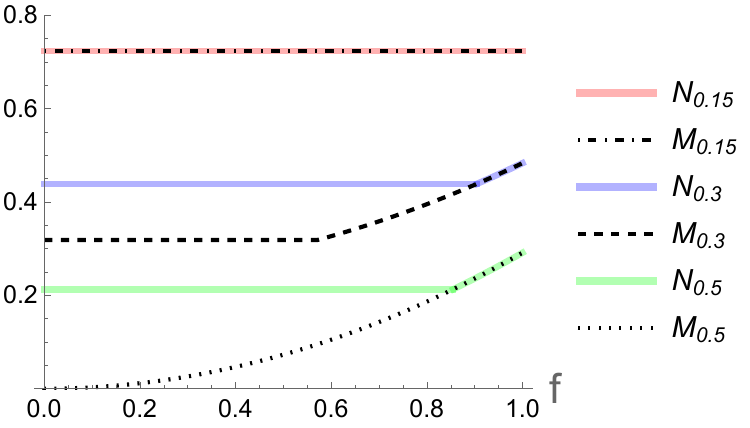}
    \caption{ORT measure $\mathcal{N}$ and metrological power $\mathcal{M}$ for a partially-coherent mixture of even and odd cat states, for different choices of the even-cat population $p=0.15,\ 0.3$ and  $0.5$, as a function of the coherence strength $f$ (see Eq.~\eqref{eq:partiallyCoherent2Cat}). The cat states have parameter $\alpha=0.5$.}
    \label{fig:partiallyCoherentMixture2CatStates}
\end{figure}

The $\alpha=0.5$ case is plotted in Fig. \ref{fig:partiallyCoherentMixture2CatStates}, for different values of the even-cat population $p$, with respect to coherence strength $f$. Unlike the previous example, here, whether the loss of coherence affects nonclassicality depends on the value of $p$. This is because, when $|\langle\hat{a}^2\rangle|\geq p(1-p)(r_{21}+r_{12})^2$, Eq.~\eqref{eq:overSqueezed2} holds, regardless of $f$, which corresponds to values of $p$ outside of the interval $(p_L,p_R)=((1-e^{-2|\alpha|^2})/2,(1+e^{-2|\alpha|^2})/2)$. For $\alpha=0.5$, $(p_L,p_R)\approx(.197,.803)$. Hence, $\mathcal{N}$ is flat with respect to $f$ in the $p=0.15$ case. Between $(p_L,p_R)$, there are critical coherence values satisfying $\eta(f_\text{crit}(p),p)=|\langle\hat{a}^2\rangle|$, above which the nonclassicality varies with $f$. However, $f_\text{crit}$ depends on $p$, attaining its lowest value $f_\text{crit}^\text{min}\approx0.855$ at $p=0.5$. Metrological power matches nonclassicality above $f_\text{crit}$. Below $f_\text{crit}$, $\mathcal{M}$'s behavior with respect to $f$ is different in the $p=0.3$ and $p=0.5$ cases, despite these both falling in the interval $(p_L,p_R)$.

\subsection{Remark on Metrological Power}
Recall that $f=1$ corresponds to a pure state, and for pure states $\mathcal{M}(\ket{\psi})=\mathcal{N}(\ket{\psi})$ always. So, metrological power and nonclassicality must approach one another as $f\rightarrow1$ (i.e., as coherence becomes maximal). Our examples show that we can obtain exact matching at partial coherence. In Fig. \ref{fig:partiallyCoherentMixture2FockStates} and \ref{fig:partiallyCoherentMixture2CatStates}, metrological power becomes equal to nonclassicality at values of $f$ below $1$, although the sufficient value may vary in individual cases. In some cases, $\mathcal{M}=\mathcal{N}$ even in maximally incoherent cases ($f=0$), as seen in Fig. \ref{fig:partiallyCoherentMixture2CatStates} and alluded to in Eq.~\eqref{eq:overSqueezed} and~\eqref{eq:M1}.

\section{Higher Rank States}\label{sec:higherRank}
In Appendix \ref{sec:appendixHighRank}, we provide a method to numerically calculate the ORT measure for higher-rank states meeting similar conditions to Assumptions 1*, 2*, 3, and 4. We give a few examples below (an additional example can be found in Appendix \ref{sec:appendixAdditionalExamples}). Importantly, one example demonstrates that higher photon number coherence coherence can sometimes yield less nonclassicality and metrological power.

We consider partially-coherent mixtures of three Fock states:
\begin{small}
\begin{equation}
\begin{split}
&\hat{\rho}=\\
&\begin{pmatrix}
p_{n+2} & f_{21}\sqrt{p_{n+2}p_{n+1}}e^{i\phi} & f_{20}\sqrt{p_{n+2}p_{n}}e^{2i\phi}\\
f_{21}\sqrt{p_{n+2}p_{n+1}}e^{-i\phi} & p_{n+1} & f_{10}\sqrt{p_{n+1}p_n}e^{i\phi}\\
f_{20}\sqrt{p_{n+2}p_{n}}e^{-2i\phi} & f_{10}\sqrt{p_{n+1}p_n}e^{-i\phi} & p_n
\end{pmatrix},
\end{split}
\end{equation}
\end{small}
where the $f$'s are between 0 and $1$ and $f_{21}^2+f_{20}^2+f_{10}^2\leq1+2f_{21}f_{20}f_{10}$ to guarantee a positive matrix. For a comprehensive treatment of the \textit{incoherent} mixture ($f_{21}=f_{10}=f_{20}=0$ case), see Ref. \cite{rogers2024quantifying}.
\begin{figure}
    \centering
     \captionsetup{justification=raggedright, singlelinecheck=false}
    \includegraphics[width=\columnwidth]{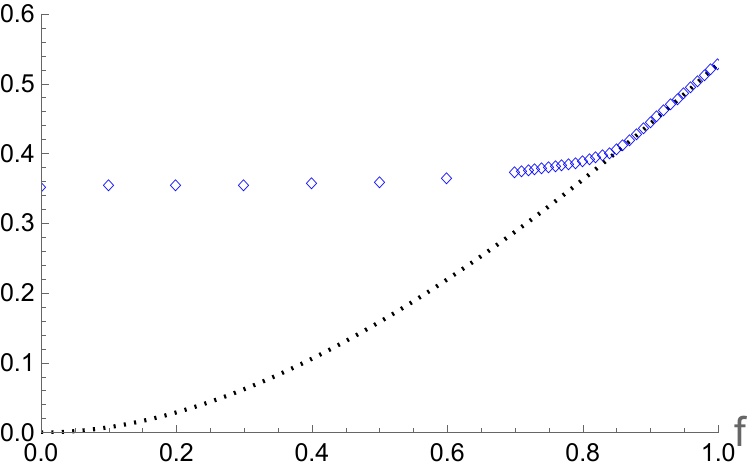}
    \caption{Numerically-calculated ORT measure $\mathcal{N}$ (blue diamonds) and exact metrological power $\mathcal{M}$ (dotted, black) for partially-coherent mixtures of $\ket{2}$, $\ket{1}$, and $\ket{0}$ with equal photon number populations $1/3$ and equal photon number coherences (proportional to $f$, see Eq.~\eqref{eq:threeMixtureExample}).}
    \label{fig:thirdMixture}
\end{figure}

Fig. \ref{fig:thirdMixture}, shows the ORT measure for the set of states:
\begin{equation}
\hat{\rho}=\frac{1}{3}\begin{pmatrix}
1 & f & f\\
f & 1 & f\\
f & f & 1
\end{pmatrix},
\label{eq:threeMixtureExample}
\end{equation}
where we have taken $0\leq f\leq1$ and the lowest photon number to be $n=0$. This means $\hat{\rho}$ in Eq.~\eqref{eq:threeMixtureExample} has eigenstates:
\begin{subequations}
\begin{align}
\ket{\psi_1}&=\frac{1}{\sqrt{3}}\left(\ket{2}+\ket{1}+\ket{0}\right)\\
\ket{\psi_2}&=\frac{1}{\sqrt{3}}\left(\ket{2}+e^{2\pi i/3}\ket{1}+e^{4\pi i/3}\ket{0}\right)\\
\ket{\psi_3}&=\frac{1}{\sqrt{3}}\left(\ket{2}+e^{4\pi i/3}\ket{1}+e^{2\pi i/3}\ket{0}\right),
\end{align}
\end{subequations}
and eigenvalues $\lambda_1=(2f+1)/3$, $\lambda_2=\lambda_3=(1-f)/3$. $\mathcal{N}$ increases monotonically in $f$, in agreement with Eq.~\eqref{eq:generalBDCBound}. There is a critical value $f_\text{crit}$, between .86 and .87, at which the derivative of $\mathcal{N}$ sharply changes and the witness term's contribution (see Eq.~\eqref{eq:ORTMixed}) becomes nonzero; there is also strong numerical evidence that above $f_\text{crit}$, $\mathcal{N}=\mathcal{M}$. At $f=1$, $\mathcal{N}=\mathcal{M}$ for certain, since $\hat{\rho}$ becomes pure.

\begin{figure}
    \centering
     \captionsetup{justification=raggedright, singlelinecheck=false}
    \includegraphics[width=\columnwidth]{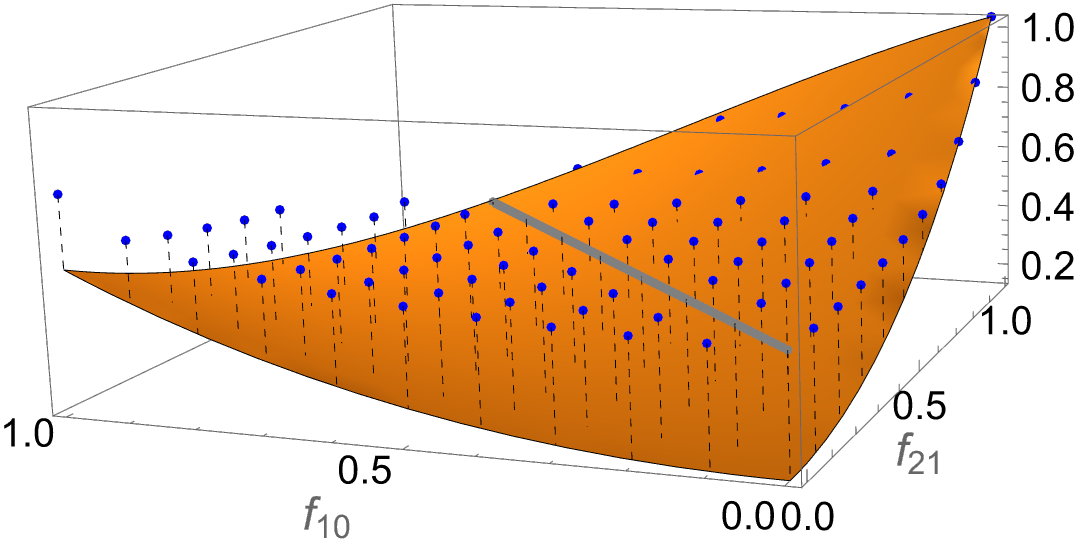}
    \caption{Numerically-calculated ORT measure (dots) and metrological power (orange surface) for the states in Eq.~\eqref{eq:threeMixtureExample3}, with $p_{n+2}=0.4$, $p_{n+1}=0.4$, $p_n=0.2$, and $n=0$. For large enough $f_{21}$, the ORT measure and metrological power agree.}
    \label{fig:valley}
\end{figure}

Another example is the set of states:
\begin{equation}
\hat{\rho}=\begin{pmatrix}
p_{n+2} & f\sqrt{p_{n+2}p_{n+1}} & 0\\
f\sqrt{p_{n+2}p_{n+1}} & p_{n+1} & f\sqrt{p_{n+1}p_{n}}\\
0 & f\sqrt{p_{n+1}p_{n}} & p_n
\end{pmatrix}.
\label{eq:threeMixtureExample2}
\end{equation}
To ensure positivity, $f\leq\frac{1}{\sqrt{2}}$. For a large class of photon number populations, these states saturate the lower bound in Eq.~\eqref{eq:noCoherenceVsWithCoherence}. In Ref. ~\cite{rogers2024quantifying}, it was established numerically, for the $f=0$ case, that if the populations satisfy:
\begin{subequations}
\begin{align}
p_{n+2}+p_{n+1}&\geq\frac{(2+n)p_{n+2}}{(1+n)p_{n+1}+(3+2n)p_{n+2}}\label{eq:populationCondition1}\\
1-p_{n+2}&\geq\frac{(1+n)(-1+p_{n+1}+p_{n+2})}{-3-2n+(1+n)p_{n+1}+(3+2n)p_{n+2}},\label{eq:populationCondition2}
\end{align}
\end{subequations}
then the ORT measure is:
\begin{equation}
\mathcal{N}(\hat{\rho}_0)=\langle\hat{n}\rangle-\left(\sqrt{p_{n+2}p_{n+1}(n+2)}+\sqrt{p_{n+1}p_{n}(n+1)}\right)^2.\label{eq:incoherentFock}
\end{equation}
For all $f$ between $0$ and $\frac{1}{\sqrt{2}}$, the decomposition of $\hat{\rho}$ in Eq.~\eqref{eq:threeMixtureExample2} into states:
\begin{equation}
\begin{split}
\ket{\phi_{\pm,\pm}}&=\sqrt{p_{n+2}}\ket{n+2}+e^{i\theta_{\pm,\pm}}\sqrt{p_{n+1}}\ket{n+1}\\&+e^{2i\theta_{\pm,\pm}}\sqrt{p_{n}}\ket{n},
\end{split}
\end{equation}
with angles $\theta_{\pm,\pm}=\pm\arccos\left({f\pm\frac{1}{2}\sqrt{2-4f^2}}\right)$ and probabilities $1/4$ each, gives the same value as $\hat{\rho}_0$ when substituted into the objective function on the right hand side of Eq.~\eqref{eq:ORTMixed}. This must be an optimal decomposition for the populations satisfying Eq.~\eqref{eq:populationCondition1}--\eqref{eq:populationCondition2}, since it saturates the lower bound in Eq.~\eqref{eq:noCoherenceVsWithCoherence}. Fig. \ref{fig:valley} shows the more general case:

\begin{equation}
\hat{\rho}=\begin{pmatrix}
p_{n+2} & f_{21}\sqrt{p_{n+2}p_{n+1}} & 0\\
f_{21}\sqrt{p_{n+2}p_{n+1}} & p_{n+1} & f_{10}\sqrt{p_{n+1}p_{n}}\\
0 & f_{10}\sqrt{p_{n+1}p_{n}} & p_n
\end{pmatrix}
\label{eq:threeMixtureExample3}
\end{equation}
with a choice of populations $p_{n+2}=0.4$, $p_{n+1}=0.4$, $p_n=0.2$. This choice satisfies Eq.~\eqref{eq:populationCondition1}--\eqref{eq:populationCondition2} and indeed  a constant valley extends from the zero-coherence case ($f_{21}=f_{10}=0$) along the $f_{21}=f_{10}$ diagonal (gray line), indicating that the ORT measure is the same value as if there were no photon number coherence (i.e., the inequality in Eq.~\eqref{eq:noCoherenceVsWithCoherence} is saturated).

In addition, Fig. \ref{fig:valley} shows that more coherence does not necessarily mean more nonclassicality and metrological power. For the purposes of this paragraph, let $(f_{21},f_{10})$ specify a state in Fig. \ref{fig:valley}. To give one example, we find that $\mathcal{N}(0.2, 0.6)\approx0.511$ and $\mathcal{M}(0.2,0.6)\approx0.233$, while $\mathcal{N}(0,0.5)\approx0.523$ and $\mathcal{M}(0,0.5)\approx0.235$. So, while $(0.2,0.6)$ has strictly larger photon number coherences than $(0,0.5)$, it has less nonclassicality and metrological power. This is possible because $(0.2,0.6)$ and $(0,0.5)$ are not connected by a bosonic dephasing channel, and thus Eq. \eqref{eq:generalBDCBound}-\eqref{eq:metrologicalPowerBDCBound} does not apply to the pair of states. As described in Appendix \ref{sec:appendixDephasing}, a bosonic dephasing channel must reduce $f_{21}$ and ${f_{10}}$ by the same factor.

\section{Conclusion}
In this work, we have calculated the ORT nonclassicality measure and metrological power for various bosonic mixed states. Predominantly, we focused on a class of rank-2 mixed states spanned by special bases (including, but not limited to, Fock states, cat states, and squeezed states). These special bases satisfy simplifying assumptions, making the ORT measure analytically solvable. By comparing, we identified regimes where the ORT measure equals metrological power. Furthermore, we found that both nonclassicality and metrological power could exhibit piecewise behavior as a function of populations and coherence. For example, a ``sudden death" phenomena is possible whereby nonclassicality or metrological power suddenly plateaus as coherence is reduced.

We especially considered the impact of coherence on nonclassicality and metrological power. We showed that a state with photon number coherence is at least as nonclassical and metrologically useful as a state with the same photon number populations but zero photon number coherence. More generally, $\mathcal{N}(\hat{\rho}')\leq\mathcal{N}(\hat{\rho})$ if there exists a bosonic dephasing operation $\mathcal{D}$ such that $\mathcal{D}(\hat{\rho})=\hat{\rho}'$. However, as we point out, the existence of such an operation must be considered carefully. Even if $\hat{\rho}'$ has strictly lower photon number coherences than $\hat{\rho}$, and the same photon number populations, it may be that no bosonic dephasing operation connects them, since bosonic dephasing cannot decrease coherences arbitrarily. By examining a partially-coherent mixture of three Fock states in section \ref{sec:higherRank}, we confirmed that lowering photon number coherences, in a manner inconsistent with bosonic dephasing, can increase nonclassicality. Thus, certain decreases to photon number coherences may require nonclassical resources to bring them about.

Lack of purity and coherence are generally perceived as undesirable, but inevitable features of quantum systems. By analyzing mixed states, we have shown how these features may impact nonclassicality in unexpected and interesting ways, with consequences for quantum optical sensors. Future work could quantify the effect of processes besides dephasing and broaden the class of calculable states via free operations.

\section*{Acknowledgements}
We thank Garrett Jepson and Luke Ellert-Beck for helpful discussions.

This research is supported by NSF Award No. 2243591. The work of Spencer Rogers is also supported by NIST Award No. 60NANB23D166.

\newpage
\appendix
\begin{widetext}
\section{Partially-coherent mixtures of Three Fock states connected by dephasing}\label{sec:appendixDephasing}
Let $\hat{\rho}$ and $\hat{\rho}'$ be partially-coherent mixtures of Fock states $\ket{n+2}$, $\ket{n+1}$, and $\ket{n}$ with equal photon number populations and $\arg[\rho'_{lm}]=\arg[\rho_{lm}]$ for all $(l,m)$. Then $\hat{\rho}'=\mathcal{D}_{\text{sym},p}(\hat{\rho})$ for some $\mathcal{D}_{\text{sym},p}$ (see Eq.~\eqref{eq:symmetricDephasing}--\eqref{eq:symmetricDephasingChannelEffect}) if and only if:
\begin{subequations}
\begin{align}
\bra{n+2}\hat{\rho}'\ket{n+1}&=x{\bra{n+2}\hat{\rho}\ket{n+1}}\label{eq:ratioOfCoherences}\\{\bra{n+1}\hat{\rho}'\ket{n}}&=x{\bra{n+1}\hat{\rho}\ket{n}}\\
{\bra{n+2}\hat{\rho}'\ket{n}}&=y{\bra{n+2}\hat{\rho}\ket{n}}\\
|x|&\leq1\\
|y|&\leq1\\
2x^2-1&\leq y,\label{eq:exampleOfCosInequality}
\end{align}
\end{subequations}
for some choice of $x$ and $y$.
Eq.~\eqref{eq:exampleOfCosInequality} utilizes the inequality $\langle\cos2\phi\rangle\geq2\langle\cos\phi\rangle^2-1$. The nearest-neighbor coherences must be reduced by the same factor $x$, which sets a lower bound on the reduction factor $y$ for the $\ket{n+2}$-$\ket{n}$ coherence. Eq.~\eqref{eq:ratioOfCoherences}-\ref{eq:exampleOfCosInequality} are very strict requirements, and will not be met for most choices of $\hat{\rho}$ and $\hat{\rho}'$.

\section{Derivation of ORT measure for Rank-2 Mixtures}\label{sec:appendixDerivation}
\subsection{Eq.~\eqref{eq:overSqueezed}--\eqref{eq:underSqueezed}}
We now explain the derivation of the ORT measure, Eq.~\eqref{eq:overSqueezed}--\eqref{eq:underSqueezed} for rank-2 mixtures $\hat{\rho}=p\ket{\psi_1}\bra{\psi_1}+(1-p)\ket{\psi_2}\bra{\psi_2}$ satisfying Assumptions 1 and 2. For completeness, we restate those assumptions here.

\textbf{Assumption 1: The two eigenstates are centered at the origin in phase space.} This means $\bra{\psi_1}\hat{a}\ket{\psi_1}=\bra{\psi_2}\hat{a}\ket{\psi_2}=0$.

\textbf{Assumption 2: The eigenstates satisfy $\arg(\bra{\psi_1}\hat{a}^2\ket{\psi_1})=\arg(\bra{\psi_2}\hat{a}^2\ket{\psi_2})=\arg(\bra{\psi_1}\hat{a}\ket{\psi_2})+\arg(\bra{\psi_2}\hat{a}\ket{\psi_1})$.} Defining $\mu\equiv\arg(\bra{\psi_1}\hat{a}^2\ket{\psi_1})/2$, it is then always possible to perform a change of basis, $\ket{\psi_1}\rightarrow e^{i\theta_1}\ket{\psi_1}$ and $\ket{\psi_2}\rightarrow e^{i\theta_2}\ket{\psi_2}$, such that:
\begin{subequations}
\begin{align}
\bra{\psi_1}\hat{a}\ket{\psi_2}&=|\bra{\psi_1}\hat{a}\ket{\psi_2}|e^{i\mu}\equiv r_{12}e^{i\mu}
\label{eq:assigningArg1}\\
\bra{\psi_2}\hat{a}\ket{\psi_1}&=|\bra{\psi_2}\hat{a}\ket{\psi_1}|e^{i\mu}\equiv r_{21}e^{i\mu}.
\label{eq:assigningArg2}
\end{align}
\end{subequations}
It is simplest to work in such a basis, so we will assume Eq.~\eqref{eq:assigningArg1}--\eqref{eq:assigningArg2} throughout the derivation without loss of generality.

Following Eq.~\eqref{eq:ORTMixed}, the task of calculating the ORT measure amounts to the minimization of:
\begin{equation}
\mathcal{L}[\{q_j,\ket{\phi_j}\}]=-\sum_jq_j|\bra{\phi_j}\hat{a}\ket{\phi_j}|^2+\left|\langle{\hat{a}^2}\rangle-\sum_jq_j\bra{\phi_j}\hat{a}\ket{\phi_j}^2\right|
\label{eq:objectiveAppendix}
\end{equation}
over all convex decompositions $\{q_j,\ket{\phi_j}\}$ of $\hat{\rho}$ (note that $\langle\hat{a}^\dagger\hat{a}\rangle=\langle\hat{n}\rangle$ in Eq.~\eqref{eq:ORTMixed} is a decomposition-independent quantity and can be simply added later). Only states in $\text{span}(\hat{\rho})$ are allowed; these have the general form $\ket{\phi_{x,\theta}}=x\ket{\psi_1}+\sqrt{1-x^2}e^{i\theta}\ket{\psi_2}$. We may use Assumptions 1 and 2 to write:
\begin{subequations}
\begin{align}
\bra{\phi_{x,\theta}}\hat{a}\ket{\phi_{x,\theta}}&=x\sqrt{1-x^2}e^{i\mu}\left(r_{21}e^{-i\theta}+r_{12}e^{i\theta}\right)\label{eq:aXTheta}
\\
\left|\bra{\phi_{x,\theta}}\hat{a}\ket{\phi_{x,\theta}}\right|^2&=x^2(1-x^2)\left(r_{21}^2+r_{12}^2+2r_{21}r_{12}\cos{2\theta}\right)
\\
\bra{\phi_{x,\theta}}\hat{a}\ket{\phi_{x,\theta}}^2&=x^2(1-x^2)e^{2i\mu}\left(r_{21}^2e^{-2i\theta}+r_{12}^2e^{2i\theta}+2r_{21}r_{12}\right)\\
\langle\hat{a}^2\rangle&=p\bra{\psi_1}\hat{a}^2\ket{\psi_1}+(1-p)\bra{\psi_2}\hat{a}^2\ket{\psi_2}=|\langle\hat{a}^2\rangle|e^{2i\mu}.\label{eq:aSquaredAppendix}
\end{align}
\end{subequations}

Then we may rewrite the objective as:
\begin{equation}
\begin{split}
\mathcal{L}[q_{x,\theta}]&=-\sum_{x,\theta}q_{x,\theta}x^2(1-x^2)\left(r_{21}^2+r_{12}^2+2r_{21}r_{12}\cos{2\theta}\right)\\
&+\left|\left|\langle{\hat{a}^2}\rangle\right|-\sum_{x,\theta}q_{x,\theta}x^2(1-x^2)\left(r_{21}^2e^{-2i\theta}+r_{12}^2e^{2i\theta}+2r_{21}r_{12}\right)\right|,
\end{split}\label{eq:objectiveBeforeThetaSymmetrization}
\end{equation}
subject to the constraints:
\begin{subequations}
\begin{align}
\sum_{x,\theta}q_{x,\theta}x^2&=p\label{eq:populationConstraintExp1}\\
\sum_{x,\theta}q_{x,\theta}x\sqrt{1-x^2}e^{i\theta}&=0\label{eq:offDiagonalConstraintExp1}\\
\sum_{x,\theta}q_{x,\theta}&=1.
\end{align}
\end{subequations}
We note that it is sufficient to optimize over distributions $q_{x,\theta}$ that are symmetric in $\theta$: $q_{x,\theta}=q_{x,-\theta}$. This is because, given any decomposition $q_{x,\theta}$ satisfying the constraints, one may always construct the symmetric distribution $\tilde{q}_{x,\theta}\equiv\left(q_{x,\theta}+q_{x,-\theta}\right)/2$ which also satisfies the constraints and produces a lower (or equal) value of the objective: $\mathcal{L}[\tilde{q}_{x,\theta}]\leq\mathcal{L}[q_{x,\theta}]$. Thus, the ultimate goal is to minimize:
\begin{equation}
\tilde{\mathcal{L}}=-\langle x^2(1-x^2)\left(r_{21}^2+r_{12}^2+2r_{21}r_{12}\cos{2\theta}\right)\rangle+\left|\left|\langle{\hat{a}^2}\rangle\right|-\langle x^2(1-x^2)\left[\left(r_{21}^2+r_{12}^2\right)\cos2\theta+2r_{21}r_{12}\right]\rangle\right|,
\label{eq:objectiveAfterThetaSymmetrization}
\end{equation}
subject to the constraints:
\begin{subequations}
\begin{align}
\langle x^2\rangle&=p\label{eq:populationConstraintCos1}\\
\langle x\sqrt{1-x^2}\cos\theta\rangle&=0\label{eq:offDiagonalConstraintCos1}\\
q_{x,\theta}&=q_{x,-\theta}\geq0.
\end{align}
\end{subequations}
Here, we are using the shorthand $\langle A\rangle\equiv\sum_{x,\theta}q_{x,\theta}A$, not to be confused with operator expectation values $\langle \hat{A}\rangle\equiv\tr\left(\hat{\rho}\hat{A}\right)$. Using Jensen's inequality (which implies $\langle x^4\rangle\geq\langle x^2\rangle^2$) and Eq.~\eqref{eq:populationConstraintCos1}, we may conclude:
\begin{subequations}
\begin{align}
\langle x^2(1-x^2)\rangle\leq p(1-p)\label{eq:constraintInequalityPop1}\\
-p(1-p)\leq\langle x^2(1-x^2)\cos2\theta\rangle\leq p(1-p).\label{eq:constraintInequalityCos1}
\end{align}
\end{subequations}
These bounds are important because $\langle x^2(1-x^2)\rangle$ and $\langle x^2(1-x)^2\cos2\theta\rangle$ constitute all decomposition-dependent quantities that occur in the objective function $\tilde{\mathcal{L}}$, Eq.~\eqref{eq:objectiveAfterThetaSymmetrization}. It follows from them that:
\begin{equation}
\langle x^2(1-x^2)\left[\left(r_{21}^2+r_{12}^2\right)\cos2\theta+2r_{21}r_{12}\right]\rangle\leq p(1-p)(r_{21}+r_{12})^2.
\end{equation}
Concerning the absolute value term in Eq.~\eqref{eq:objectiveAfterThetaSymmetrization}, there are two possibilities: either $|\langle \hat{a}^2\rangle|\geq p(1-p)(r_{21}+r_{12})^2$ or $|\langle \hat{a}^2\rangle|< p(1-p)(r_{21}+r_{12})^2$.

In the first case, we minimize:
\begin{subequations}
\begin{align}
\tilde{\mathcal{L}}_+&=\left|\langle{\hat{a}^2}\rangle\right|-\langle x^2(1-x^2)\left[\left(r_{21}^2+r_{12}^2\right)+2r_{21}r_{12}\right](1+\cos2\theta)\rangle
\label{eq:reduceL+}\\
&\geq \left|\langle{\hat{a}^2}\rangle\right|-2p(1-p)(r_{21}+r_{12})^2
\label{eq:lowerBoundL+}\\
|\langle\hat{a}^2\rangle|&\geq\langle x^2(1-x^2)\left[\left(r_{21}^2+r_{12}^2\right)\cos2\theta+2r_{21}r_{12}\right]\rangle.
\label{eq:assumptionToReduceL+}
\end{align}
\end{subequations}
The lower bound on $\tilde{\mathcal{L}}_+$, Eq.~\eqref{eq:lowerBoundL+} utilizes the inequalities Eq.~\eqref{eq:constraintInequalityPop1}--\eqref{eq:constraintInequalityCos1}. If $|\langle \hat{a}^2\rangle|\geq p(1-p)(r_{21}+r_{12})^2$, this bound can be saturated via the decomposition into states $\sqrt{p}\ket{\psi_1}\pm\sqrt{1-p}\ket{\psi_2}$ with probabilities $\frac{1}{2}$. For this decomposition:
\begin{subequations}
\begin{align}
\langle x^2(1-x^2)\rangle&=p(1-p)\label{eq:overSqueezedDecompPop}\\
\langle x^2(1-x^2)\cos2\theta\rangle&=p(1-p).\label{eq:overSqueezedDecompCos}
\end{align}
\end{subequations}
Such a decomposition is optimal in this case, and Eq.~\eqref{eq:overSqueezed} is obtained by adding $\langle\hat{n}\rangle$ to the right hand side of Eq.~\eqref{eq:lowerBoundL+}. Eq.~\eqref{eq:assumptionToReduceL+} (included above for reference) is the general condition that allows one to derive Eq.~\eqref{eq:reduceL+} from Eq.~\eqref{eq:objectiveAfterThetaSymmetrization}; it is always met if $|\langle \hat{a}^2\rangle|\geq p(1-p)(r_{21}+r_{12})^2$, regardless of the decomposition $q_{x,\theta}$.

In the second case, $|\langle \hat{a}^2\rangle|< p(1-p)(r_{21}+r_{12})^2$, we must account for the fact that the absolute value in Eq.~\eqref{eq:objectiveAfterThetaSymmetrization} may multiply what is inside by either $1$ or $-1$. Thus we must find $\min\left(\min(\tilde{\mathcal{L}}_+),\min(\tilde{\mathcal{L}}_-)\right)$, where:
\begin{subequations}
\begin{align}
\tilde{\mathcal{L}}_-&=-|\langle\hat{a}^2\rangle|+\langle x^2(1-x^2)[r_{21}^2+r_{12}^2-2r_{21}r_{12}]\left(-1+\cos2\theta\right)\rangle
\label{eq:reduceL-}\\
|\langle\hat{a}^2\rangle|&\leq\langle x^2(1-x^2)\left[\left(r_{21}^2+r_{12}^2\right)\cos2\theta+2r_{21}r_{12}\right]\rangle.
\label{eq:assumptionToReduceL-}
\end{align}
\end{subequations}
Eq.~\eqref{eq:assumptionToReduceL-} (included above for reference) is the general condition that allows one to derive Eq.~\eqref{eq:reduceL-} from Eq.~\eqref{eq:objectiveAfterThetaSymmetrization}; if $|\langle \hat{a}^2\rangle|< p(1-p)(r_{21}+r_{12})^2$, whether it is met depends on the decomposition $q_{x,\theta}$. It turns out that $\min(\tilde{\mathcal{L}}_+)=\min(\tilde{\mathcal{L}}_-)$, when $|\langle \hat{a}^2\rangle|< p(1-p)(r_{21}+r_{12})^2$. Let us start with $\min(\tilde{\mathcal{L}}_-)$. Rewriting Eq.~\eqref{eq:assumptionToReduceL-} as:
\begin{equation}
\frac{|\langle\hat{a}^2\rangle|-2r_{21}r_{12}\langle x^2(1-x^2)\rangle}{r_{21}^2+r_{12}^2}\leq\langle x^2(1-x^2)\cos2\theta\rangle,
\end{equation}
and combining with Eq.~\eqref{eq:reduceL-} gives:
\begin{equation}
\begin{split}
\tilde{\mathcal{L}}_-&\geq\frac{-1}{r_{21}^2+r_{12}^2}\left(2r_{21}r_{12}|\langle\hat{a}^2\rangle|+(r_{21}^2-r_{12}^2)^2\langle x^2(1-x^2)\rangle\right)\\
&\geq\frac{-1}{r_{21}^2+r_{12}^2}\left(2r_{21}r_{12}|\langle\hat{a}^2\rangle|+(r_{21}^2-r_{12}^2)^2p(1-p)\right).\label{eq:lowerBoundL-}
\end{split}
\end{equation}
The lower bound on the second line is obtained using Eq.~\eqref{eq:constraintInequalityPop1} and may be saturated via a decomposition $\{q_\theta,\ket{\psi_\theta}=\sqrt{p}\ket{\psi_1}+\sqrt{1-p}e^{i\theta}\ket{\psi_2}\}$ satisfying:
\begin{equation}
\bar{X}\equiv\frac{|\langle\hat{a}^2\rangle|-2r_{21}r_{12}p(1-p)}{p(1-p)(r_{21}^2+r_{12}^2)}=\langle\cos2\theta\rangle.\label{eq:barXAppendix}
\end{equation}
Since $0\leq|\langle\hat{a}^2\rangle|< p(1-p)(r_{21}+r_{12})^2$ (by assumption) and $2r_{21}r_{12}\leq r_{21}^2+r_{12}^2$ always, we have $-1\leq\bar{X}<1$. It is always possible to obtain the necessary $\langle\cos2\theta\rangle$ using weights: $q_0=q_\pi=(\bar{X}+1)/4$ and $q_{\pi/2}=q_{-\pi/2}=(1-\bar{X})/4$. For this decomposition:
\begin{subequations}
\begin{align}
\langle x^2(1-x^2)\rangle&=p(1-p)\label{eq:underSqueezedDecompPopA}\\
\langle x^2(1-x^2)\cos2\theta\rangle&=p(1-p)\bar{X}.\label{eq:underSqueezedDecompCosA}
\end{align}
\end{subequations}
Such a decomposition is optimal in this case (as we will confirm by comparing against $\min(\mathcal{L}_+)$ below), and Eq.~\eqref{eq:underSqueezed} is obtained by adding $\langle\hat{n}\rangle$ to the right hand side of Eq.~\eqref{eq:lowerBoundL-}. Next let us consider $\tilde{\mathcal{L}}_+$ (assuming $|\langle\hat{a}^2\rangle|< p(1-p)(r_{21}+r_{12})^2$). Rewriting Eq.~\eqref{eq:assumptionToReduceL+} as:
\begin{equation}
\frac{|\langle\hat{a}^2\rangle|-2r_{21}r_{12}\langle x^2(1-x^2)\rangle}{r_{21}^2+r_{12}^2}\geq\langle x^2(1-x^2)\cos2\theta\rangle,\label{eq:assumptionToReduceL+Equivalent}
\end{equation}
and combining with Eq.~\eqref{eq:reduceL+} gives:
\begin{equation}
\begin{split}
\tilde{\mathcal{L}}_+&\geq\frac{-1}{r_{21}^2+r_{12}^2}\left(2r_{21}r_{12}|\langle\hat{a}^2\rangle|+(r_{21}^2-r_{12}^2)^2\langle x^2(1-x^2)\rangle\right)\\
&\geq\frac{-1}{r_{21}^2+r_{12}^2}\left(2r_{21}r_{12}|\langle\hat{a}^2\rangle|+(r_{21}^2-r_{12}^2)^2p(1-p)\right).
\end{split}
\end{equation}
The lower bound on the second line is obtained using Eq.~\eqref{eq:constraintInequalityPop1}. This is the same lower bound obtained for $\tilde{\mathcal{L}}_-$ and may be saturated using the same decomposition. This optimal decomposition essentially zeroes out the absolute value in Eq.~\eqref{eq:objectiveAfterThetaSymmetrization}, and we have $\min(\tilde{\mathcal{L}}_+)=\min(\tilde{\mathcal{L}}_-)$.


In summary, we have:
\begin{subequations}
\begin{align}
\min\tilde{\mathcal{L}}&=|\langle\hat{a}^2\rangle|-2p(1-p)(r_{21}+r_{12})^2 &  |\langle\hat{a}^2\rangle|&\geq p(1-p)(r_{21}+r_{12})^2\\
\min\tilde{\mathcal{L}}&=\frac{-1}{r_{21}^2+r_{12}^2}\left(2r_{21}r_{12}|\langle\hat{a}^2\rangle|+(r_{21}^2-r_{12}^2)^2p(1-p)\right)&  |\langle\hat{a}^2\rangle|&< p(1-p)(r_{21}+r_{12})^2.
\end{align}
\end{subequations}
Adding $\langle\hat{n}\rangle$ gives Eq.~\eqref{eq:overSqueezed}--\eqref{eq:underSqueezed}.

\subsection{Eq.~\eqref{eq:overSqueezed2}--\eqref{eq:underSqueezedB}}
We now explain the derivation of the ORT measure, Eq.~\eqref{eq:overSqueezed2}--\eqref{eq:underSqueezedB} for rank-2 mixtures $\hat{\rho}=p\ket{\psi_1}\bra{\psi_1}+(1-p)\ket{\psi_2}\bra{\psi_2}+fe^{i\chi}\sqrt{p(1-p)}\ket{\psi_1}\bra{\psi_2}+fe^{-i\chi}\sqrt{p(1-p)}\ket{\psi_2}\bra{\psi_1}$ 
satisfying Assumptions 1*, 2*, 3, and 4. For completeness, we restate those assumptions here. Note that $0\leq f\leq1$ and that we refer to $\ket{\psi_1}$ and $\ket{\psi_2}$ as the \textit{special basis states} (as opposed to eigenstates).

\textbf{Assumption 1*: The special basis states are centered at the origin in phase space.} By this we mean that $\bra{\psi_1}\hat{a}\ket{\psi_1}=\bra{\psi_2}\hat{a}\ket{\psi_2}=0$.

\textbf{Assumption 2*: The special basis states satisfy $\arg(\bra{\psi_1}\hat{a}^2\ket{\psi_1})=\arg(\bra{\psi_2}\hat{a}^2\ket{\psi_2})=\arg(\bra{\psi_1}\hat{a}\ket{\psi_2})+\arg(\bra{\psi_2}\hat{a}\ket{\psi_1})$.} Defining $\mu\equiv\arg(\bra{\psi_1}\hat{a}^2\ket{\psi_1})/2$, it is then always possible to perform a change of basis, $\ket{\psi_1}\rightarrow e^{i\theta_1}\ket{\psi_1}$ and $\ket{\psi_2}\rightarrow e^{i\theta_2}\ket{\psi_2}$, such that:
\begin{subequations}
\begin{align}
\bra{\psi_1}\hat{a}\ket{\psi_2}&=|\bra{\psi_1}\hat{a}\ket{\psi_2}|e^{i\mu}\equiv r_{12}e^{i\mu}
\label{eq:assigningArg1Again}\\
\bra{\psi_2}\hat{a}\ket{\psi_1}&=|\bra{\psi_2}\hat{a}\ket{\psi_1}|e^{i\mu}\equiv r_{21}e^{i\mu}.
\label{eq:assigningArg2Again}
\end{align}
\end{subequations}
We will call a basis satisfying Eq.~\eqref{eq:assigningArg1Again}--\eqref{eq:assigningArg2Again} a \textit{convenient special basis} and make use of it later.

\textbf{Assumption 3: The phase of the coherence satisfies} $2\chi\mod2\pi=\arg(\bra{\psi_1}\hat{a}\ket{\psi_2})-\arg(\bra{\psi_2}\hat{a}\ket{\psi_1})$. In a convenient special basis (see Assumption 2*), Assumption 3 is the condition that $\chi=0$ or $\pi$.

\textbf{Assumption 4: The special basis states and $\chi$ satisfy $e^{-i\chi}\bra{\psi_1}\hat{a}^2\ket{\psi_2}+e^{i\chi}\bra{\psi_2}\hat{a}^2\ket{\psi_1}=0$.} This makes $\langle\hat{a}^2\rangle$ invariant under $f$. In a convenient special basis (see Assumption 2*), Assumption 4 is the condition that $\bra{\psi_1}\hat{a}^2\ket{\psi_2}+\bra{\psi_2}\hat{a}^2\ket{\psi_1}=0$.

Given a state $\hat{\rho}$ satisfying these assumptions, it is simplest to derive Eq.~\eqref{eq:overSqueezed2}--\eqref{eq:underSqueezedB} in the convenient special basis where Eq.~\eqref{eq:assigningArg1Again}--\eqref{eq:assigningArg2Again} are satisfied, and $\chi=0$ such that:
\begin{equation}
\begin{split}
\hat{\rho}&=p\ket{\psi_1}\bra{\psi_1}+f\sqrt{p(1-p)}\ket{\psi_1}\bra{\psi_2}\\
&+f\sqrt{p(1-p)}\ket{\psi_2}\bra{\psi_1}+(1-p)\ket{\psi_2}\bra{\psi_2}.
\end{split}
\label{eq:partiallyCoherentMixtureAppendix}
\end{equation}
Without loss of generality, we work in such a basis throughout this derivation. For those interested in mapping between Eq.~\eqref{eq:partiallyCoherentMixtureAppendix} and Eq.~\eqref{eq:partiallyCoherentMixture}, note that these may be related by: $\ket{\psi_1^{\text{convenient}}}=\ket{\psi_1^{\text{old}}}$, $\ket{\psi_2^{\text{convenient}}}=e^{-i\chi}\ket{\psi_2^{\text{old}}}$ where ``convenient" and ``old" correspond to Eq.~\eqref{eq:partiallyCoherentMixtureAppendix} and Eq.~\eqref{eq:partiallyCoherentMixture} respectively.

The derivation follows many of the same steps as that of Section \ref{sec:Rank2NoCoherence}, and the reader should be well-acquainted with that derivation before proceeding. As before, to calculate the ORT measure, we must find a probability distribution $q_{x,\theta}$ of states $\ket{\phi_{x,\theta}}=x\ket{\psi_1}+\sqrt{1-x^2}e^{i\theta}\ket{\psi_2}$ that minimizes the objective function of Eq.~\eqref{eq:objectiveAppendix}. Eq.~\eqref{eq:aXTheta}--\eqref{eq:objectiveBeforeThetaSymmetrization} remain valid due to assumptions 1*, 2* and 4. Assumption 4 plays a key role in maintaining the validity of Eq.~\eqref{eq:aSquaredAppendix}, and by extension Eq.~\eqref{eq:objectiveBeforeThetaSymmetrization}. However, due to the off-diagonal elements in $\hat{\rho}$, we have different constraints on $q_{x,\theta}$:
\begin{subequations}
\begin{align}
\sum_{x,\theta}q_{x,\theta}x^2&=p\label{eq:populationConstraintExp2}\\
\sum_{x,\theta}q_{x,\theta}x\sqrt{1-x^2}e^{i\theta}&=f\sqrt{p(1-p)}\label{eq:offDiagonalConstraintExp2}\\
\sum_{x,\theta}q_{x,\theta}&=1.
\end{align}
\end{subequations}
the difference versus Eq.~\eqref{eq:populationConstraintExp1}--\eqref{eq:offDiagonalConstraintExp1} being that the right hand side of Eq.~\eqref{eq:offDiagonalConstraintExp2} is nonzero in general. It is worth noting also that the right hand side of Eq.~\eqref{eq:offDiagonalConstraintExp2} is real (as opposed to a complex), which is related to Assumption 3 and the use of the convenient special basis.

As in the $f=0$ case, it is sufficient to optimize over distributions $q_{x,\theta}$ that are symmetric in $\theta$: $q_{x,\theta}=q_{x,-\theta}$. This is because, given any decomposition $q_{x,\theta}$ satisfying the constraints, one may always construct the symmetric distribution $\tilde{q}_{x,\theta}\equiv\left(q_{x,\theta}+q_{x,-\theta}\right)/2$ which also satisfies the constraints and produces a lower (or equal) value of the objective: $\mathcal{L}[\tilde{q}_{x,\theta}]\leq\mathcal{L}[q_{x,\theta}]$. Thus, the ultimate goal is to minimize $\tilde{\mathcal{L}}$ in Eq.~\eqref{eq:objectiveAfterThetaSymmetrization},
subject to the constraints:
\begin{subequations}
\begin{align}
\langle x^2\rangle&=p\label{eq:populationConstraintCos2}\\
\langle x\sqrt{1-x^2}\cos\theta\rangle&=f\sqrt{p(1-p)}\label{eq:offDiagonalConstraintCos2}\\
q_{x,\theta}&=q_{x,-\theta}\geq0.
\end{align}
\end{subequations}
Here, we are using the shorthand $\langle A\rangle\equiv\sum_{x,\theta}q_{x,\theta}A$, not to be confused with operator expectation values $\langle \hat{A}\rangle\equiv\tr\left(\hat{\rho}\hat{A}\right)$. Importantly, Eq.~\eqref{eq:offDiagonalConstraintCos2} is more restrictive than the analogous constraint in the $f=0$ case, Eq.~\eqref{eq:offDiagonalConstraintCos1}. By Jensen's inequality, $\langle x^2(1-x^2)\cos^2\theta\rangle\geq\langle x\sqrt{1-x^2}\cos\theta\rangle^2$. Using the trigonometric identity $\cos^2\theta=(1+\cos2\theta)/2$, we have $\langle x^2(1-x^2)\cos2\theta\rangle\geq2\langle x\sqrt{1-x^2}\cos\theta\rangle^2-\langle x^2(1-x^2)\rangle$. Jensen's inequality, when combined with Eq.~\eqref{eq:populationConstraintCos2}, also gives $\langle x^2(1-x^2)\rangle\leq p(1-p)$. Thus we ultimately obtain:
\begin{subequations}
\begin{align}
\langle x^2(1-x^2)\rangle&\leq p(1-p)\label{eq:constraintInequalityPop2}\\
(2f^2-1)p(1-p)\leq\langle x^2(1-x^2)\cos2\theta\rangle&\leq p(1-p).\label{eq:constraintInequalityCos2}
\end{align}
\end{subequations}
These bounds are important because $\langle x^2(1-x^2)\rangle$ and $\langle x^2(1-x^2)\cos2\theta\rangle$ constitute all decomposition-dependent quantities that occur in the objective function $\tilde{\mathcal{L}}$, Eq.~\eqref{eq:objectiveAfterThetaSymmetrization}. Derived from the constraint that the decomposition reproduce the original state $\hat{\rho}$, they are more restrictive than those of the $f=0$ case, Eq.~\eqref{eq:constraintInequalityPop1}--\eqref{eq:constraintInequalityCos1}, since $2f^2-1>-1$ when $f\neq0$. This is the source of discrepancy between the ORT measure in the $f=0$ case, Eq.~\eqref{eq:overSqueezed}--\eqref{eq:underSqueezed}, and the ORT measure in the $f\neq0$ case, Eq.~\eqref{eq:overSqueezed2}--\eqref{eq:underSqueezedB}. Actually, Eq.~\eqref{eq:overSqueezed}--\eqref{eq:underSqueezed} represents a subset of the cases covered by Eq.~\eqref{eq:overSqueezed2}--\eqref{eq:underSqueezedA}, and there is no contradiction between them (note that $\eta(f,p)$ in Eq.~\eqref{eq:boundingEta} satisfies $\eta(0,p)\leq0\leq|\langle\hat{a}^2\rangle|$). $f\neq0$ brings further considerations into account, however, leading to Eq.~\eqref{eq:underSqueezedB}.

First we show that, when $|\langle\hat{a}^2\rangle|\geq p(1-p)(r_{21}+r_{12})^2$, we may obtain the same minimum as represented in Eq.~\eqref{eq:overSqueezed}, hence Eq.~\eqref{eq:overSqueezed2}. Recall from Section \ref{sec:Rank2NoCoherence} that Eq.~\eqref{eq:overSqueezed} utilizes a decomposition satisfying $\langle x^2(1-x^2)\rangle=\langle x^2(1-x^2)\cos2\theta\rangle=p(1-p)$, which is Eq.~\eqref{eq:overSqueezedDecompPop}--\eqref{eq:overSqueezedDecompCos}. These expectation values are compatible with Eq.~\eqref{eq:constraintInequalityPop2}--\eqref{eq:constraintInequalityCos2}. A satisfactory (optimal) decomposition of $\hat{\rho}$ is into states $\sqrt{p}\ket{\psi_1}\pm\sqrt{1-p}\ket{\psi_2}$ with probabilities $(1\pm f)/2$.

Next, for the $|\langle\hat{a}^2\rangle|< p(1-p)(r_{21}+r_{12})^2$ case, we show when it is possible to obtain the same minimum as represented in Eq.~\eqref{eq:underSqueezed}, hence Eq.~\eqref{eq:underSqueezedA}. Recall from Section \ref{sec:Rank2NoCoherence} that Eq.~\eqref{eq:underSqueezed} utilizes a decomposition satisfying $\langle x^2(1-x^2)\rangle=p(1-p)$ and $\langle x^2(1-x^2)\cos2\theta\rangle=p(1-p)\bar{X}$ (these are Eq.~\eqref{eq:underSqueezedDecompPopA}--\eqref{eq:underSqueezedDecompCosA}), where $\bar{X}$ is given in Eq.~\eqref{eq:barXAppendix}. Since $-1\leq\bar{X}<1$, these expectation values are compatible with Eq.~\eqref{eq:constraintInequalityPop2}--\eqref{eq:constraintInequalityCos2} when $(2f^2-1)\leq\bar{X}$, i.e.:
\begin{equation}
|\langle\hat{a}^2\rangle|\geq p(1-p)[(2f^2-1)(r_{21}^2+r_{12}^2)+2r_{21}r_{12}].\label{eq:aSquared>Eta}
\end{equation}
A satisfactory (optimal) decomposition of $\hat{\rho}$ is into states $\sqrt{p}\ket{\psi_1}+e^{i\theta_{\pm,\pm}}\sqrt{1-p}\ket{\psi_2}$, where $\theta_{\pm,\pm}=\pm\arccos({f\pm\sqrt{[\bar{X}-(2f^2-1)]/2}})$, with weights $q_{\pm,\pm}=1/4$ each. The angles $\theta_{\pm,\pm}$ angles are real because $(2f^2-1)\leq\bar{X}$. This explains Eq.~\eqref{eq:underSqueezedA}. Note that the main text defines $\eta(f,p)=p(1-p)[(2f^2-1)(r_{21}^2+r_{12}^2)+2r_{21}r_{12}]$ in Eq.~\eqref{eq:boundingEta} and writes Eq.~\eqref{eq:aSquared>Eta} as $|\langle\hat{a}^2\rangle|\geq\eta(f,p)$.

Lastly, we address the case where $|\langle\hat{a}^2\rangle|<p(1-p)[(2f^2-1)(r_{21}^2+r_{12}^2)+2r_{21}r_{12}]\leq p(1-p)(r_{21}+r_{12})^2$. This case requires nonzero $f$ ($|\langle\hat{a}^2\rangle|$ is positive semidefinite), and it is impossible to obtain the same minima as in the $f=0$ cases. Thus we must find new minima. Since the absolute value in $\tilde{\mathcal{L}}$ may either multiply what is inside by $\pm1$ in this regime, we must find $\min\left(\min(\tilde{\mathcal{L}}_+),\min(\tilde{\mathcal{L}}_-)\right)$, where $\tilde{\mathcal{L}}_+$ and $\tilde{\mathcal{L}}_-$ are given in Eq.~\eqref{eq:reduceL+} and Eq.~\eqref{eq:reduceL-}. It will turn out that $\min(\tilde{\mathcal{L}}_-)\leq\min(\tilde{\mathcal{L}}+)$. Let us start with $\tilde{\mathcal{L}}_-$. Combining Eq.~\eqref{eq:constraintInequalityPop2}--\eqref{eq:constraintInequalityCos2} and Eq.~\eqref{eq:reduceL-}, one obtains:
\begin{equation}
\tilde{\mathcal{L}}_-\geq-|\langle\hat{a}^2\rangle|+(2f^2-2)p(1-p)(r_{21}-r_{12})^2=\min(\tilde{\mathcal{L}}_-).\label{eq:lowerBoundL-f}
\end{equation}
This bound may be saturated by the decomposition into states $\sqrt{p}\ket{\psi_1}+e^{\pm i\theta_f}\sqrt{1-p}\ket{\psi_2}$, where $\theta_f=\arccos(f)$, with probabilities $\frac{1}{2}$ each. For this decomposition:
\begin{subequations}
\begin{align}
\langle x^2(1-x^2)\rangle&= p(1-p)\label{eq:underSqueezedDecompPopB}\\
\langle x^2(1-x^2)\cos2\theta\rangle&=(2f^2-1)p(1-p).\label{eq:underSqueezedDecompCosB}
\end{align}
\end{subequations}
Importantly, these values do not contradict Eq.~\eqref{eq:assumptionToReduceL-}, since $|\langle\hat{a}^2\rangle|<p(1-p)[(2f^2-1)(r_{21}^2+r_{12}^2)+2r_{21}r_{12}$ is true by assumption. This decomposition is optimal, and Eq.~\eqref{eq:underSqueezedB} is obtained by adding $\langle\hat{n}\rangle$ to the right hand side of Eq.~\eqref{eq:lowerBoundL-f}. However, to demonstrate optimality we still must consider $\min(\tilde{\mathcal{L}}_+)$. Since assumption Eq.~\eqref{eq:assumptionToReduceL+} and its equivalent Eq.~\eqref{eq:assumptionToReduceL+Equivalent} underlie $\tilde{\mathcal{L}}_+$, we have (combining Eq.~\eqref{eq:constraintInequalityCos2} and~\eqref{eq:assumptionToReduceL+Equivalent}):
\begin{equation}
\begin{split}
\frac{|\langle\hat{a}^2\rangle|-2r_{21}r_{12}\langle x^2(1-x^2)\rangle}{r_{21}^2+r_{12}^2}\geq(2f^2-1)p(1-p),\label{eq:willuseshortly}
\end{split}
\end{equation}
which we will use shortly. From Eq.~\eqref{eq:reduceL+}, the decomposition-dependent part of $\tilde{\mathcal{L}}_+$ is proportional to $-\langle x^2(1-x^2)(1+\cos2\theta)\rangle$. Thus, to minimize, $\tilde{\mathcal{L}}_+$ it serves to upper bound $\langle x^2(1-x^2)(1+\cos2\theta)\rangle$. Using Eq.~\eqref{eq:assumptionToReduceL+Equivalent} followed by Eq.~\eqref{eq:willuseshortly}, we get:
\begin{equation}
\begin{split}
\langle x^2(1-x^2)(1+\cos2\theta)\rangle&\leq\frac{|\langle\hat{a}^2\rangle|+(r_{21}-r_{12})^2\langle x^2(1-x^2)\rangle}{r_{21}^2+r_{12}^2},\\
&\leq\frac{1}{2r_{21}r_{12}}\left(|\langle\hat{a}^2\rangle|-(2f^2-1)p(1-p)(r_{21}-r_{12})^2)\right).
\end{split}
\end{equation}
Thus we have (referring back to Eq.~\eqref{eq:reduceL+}):
\begin{equation}
\tilde{\mathcal{L}}_+\geq\min(\tilde{\mathcal{L}}_+)\geq\frac{1}{2r_{21}r_{12}}\left(-(r_{21}+r_{12})^2|\langle\hat{a}^2\rangle|+(2f^2-1)p(1-p)(r_{21}-r_{12})^2(r_{21}+r_{12})^2\right).
\end{equation}
\begin{equation}
\begin{split}
&\min(\tilde{\mathcal{L}}_-)-\min(\tilde{\mathcal{L}}_+)\leq\\
&-|\langle\hat{a}^2\rangle|+(2f^2-2)p(1-p)(r_{21}-r_{12})^2+\frac{1}{2r_{21}r_{12}}\left((r_{21}+r_{12})^2|\langle\hat{a}^2\rangle|-(2f^2-1)p(1-p)(r_{21}-r_{12})^2(r_{21}+r_{12})^2\right)\\
&<0.
\end{split}
\end{equation}
To get the last line of the above inequality, one must use the given assumption $|\langle\hat{a}^2\rangle|<p(1-p)[(2f^2-1)(r_{21}^2+r_{12}^2)+2r_{21}r_{12}$. Thus $\min(\tilde{\mathcal{L}})=\min(\tilde{\mathcal{L}}_-)$ in this case, and Eq.~\eqref{eq:underSqueezedB} is correct.

\section{Methodology for Higher-Rank States}\label{sec:appendixHighRank}
Here we explain how we calculated the ORT measure of mixed states with rank greater than two, in the main text (we also used this method to verify our results about rank-2 states). We start with a mixed state:
\begin{equation}
\hat{\rho}=\sum_{j=1}^Jp_j\ket{\psi_j}\bra{\psi_j}+\sum_{k\neq j}\sum_{j=1}^Jf_{jk}\ket{\psi_j}\bra{\psi_k},
\end{equation}
where, for simplicity, $f_{jk}\geq0$ are real ($f_{jk}=f_{jk}^*=f_{kj}$) and $\hat{\rho}$ is ``full-rank" (has $J$ positive, nonzero eigenvalues). For our approach to work, we require certain conditions (similar to our assumptions about rank-2 states):

\begin{itemize}

\item{The special basis states satisfy $\bra{\psi_j}\hat{a}\ket{\psi_j}=0$.}

\item{The special basis states satisfy $\bra{\psi_j}\hat{a}\ket{\psi_k}=|\bra{\psi_j}\hat{a}\ket{\psi_k}|e^{i\mu}\equiv r_{jk}e^{i\mu}$. $\mu$ is independent of the pair $(j,k)$ and is also referenced in the next assumption. Note that $r_{jk}\neq r_{kj}$, in general.}

\item{$\langle\hat{a}^2\rangle=|\langle\hat{a}^2\rangle|e^{2i\mu}$. In particular we consider cases where $\bra{\psi_j}\hat{a}^2\ket{\psi_j}=|\bra{\psi_j}\hat{a}^2\ket{\psi_j}|e^{2i\mu}$ for all $j$ and, \textbf{if the $f_{jk}\neq0$,} $\bra{\psi_j}\hat{a}^2\ket{\psi_k}=|\bra{\psi_j}\hat{a}^2\ket{\psi_k}|e^{2i\mu}$ for all ordered pairs $(j,k)$. This way, we may freely choose the populations $p_j$ (and, if allowing for nonzero values, the coherences $f_{jk}$) without leaving the $\langle\hat{a}^2\rangle=|\langle\hat{a}^2\rangle|e^{2i\mu}$ regime.}
\end{itemize}

Following Eq.~\eqref{eq:ORTMixed}, the task of calculating the ORT measure amounts to the minimization of:
\begin{equation}
\mathcal{L}[\{q_j,\ket{\phi_j}\}]=-\sum_jq_j|\bra{\phi_j}\hat{a}\ket{\phi_j}|^2+\left|\langle{\hat{a}^2}\rangle-\sum_jq_j\bra{\phi_j}\hat{a}\ket{\phi_j}^2\right|
\end{equation}
over all convex decompositions $\{q_j,\ket{\phi_j}\}$ of $\hat{\rho}$ (note that $\langle\hat{a}^\dagger\hat{a}\rangle=\langle\hat{n}\rangle$ in Eq.~\eqref{eq:ORTMixed} is a decomposition-independent quantity and can be simply added later). Only states in $\text{span}(\hat{\rho})$ are allowed; these have the general form $\ket{\phi_{\vec{x},\vec{\theta}}}=\sum_jx_je^{i\theta_j}$, where $x_j\geq0$ and $\sum_jx_j^2=1$ (we may also take $\theta_1=0$ without loss of generality). Given our bulleted assumptions, we may write:
\begin{subequations}
\begin{align}
\bra{\phi_{\vec{x},\vec{\theta}}}\hat{a}\ket{\phi_{\vec{x},\vec{\theta}}}&=e^{i\mu}\sum_{(j,k)}(1-\delta_{jk})x_jx_ke^{i(\theta_j-\theta_k)}r_{kj}
\\
|\bra{\phi_{\vec{x},\vec{\theta}}}\hat{a}\ket{\phi_{\vec{x},\vec{\theta}}}|^2&=\sum_{(j,k)}(1-\delta_{jk})x_j^2x_k^2r_{kj}^2+\sum_{(l,m)}\sum_{(j,k)}(1-\delta_{lm})(1-\delta_{jk})(1-\delta_{jl}\delta_{km})x_jx_kx_lx_mr_{kj}r_{ml}\cos(\theta_j+\theta_m-\theta_k-\theta_l)
\\
\bra{\phi_{\vec{x},\vec{\theta}}}\hat{a}\ket{\phi_{\vec{x},\vec{\theta}}}^2&=e^{2i\mu}\left(\sum_{(j,k)}(1-\delta_{jk})x_j^2x_k^2r_{kj}r_{jk}+\sum_{(l,m)}\sum_{(j,k)}(1-\delta_{lm})(1-\delta_{jk})(1-\delta_{jm}\delta_{kl})x_jx_kx_lx_mr_{kj}r_{ml}e^{i(\theta_j+\theta_l-\theta_k-\theta_m)}\right)
\end{align}
\end{subequations}

We may rewrite the objective as:
\begin{equation}
\mathcal{L}[q_{\vec{x},\vec{\theta}}]=-\sum_{\vec{x},\vec{\theta}}q_{\vec{x},\vec{\theta}}|\bra{\phi_{\vec{x},\vec{\theta}}}\hat{a}\ket{\phi_{\vec{x},\vec{\theta}}}|^2+\left|\left|\langle{\hat{a}^2}\rangle\right|e^{2i\mu}-\sum_{\vec{x},\vec{\theta}}q_{\vec{x},\vec{\theta}}\bra{\phi_{\vec{x},\vec{\theta}}}\hat{a}\ket{\phi_{\vec{x},\vec{\theta}}}^2\right|,
\end{equation}
subject to the constraints:
\begin{subequations}
\begin{align}
\sum_{\vec{x},\vec{\theta}}q_{\vec{x},\vec{\theta}}x_j^2&=p_j \hspace{1cm} &\text{for all } j\\
\sum_{\vec{x},\vec{\theta}}q_{\vec{x},\vec{\theta}}x_jx_ke^{i(\theta_j-\theta_k)}&=f_{kj}=f_{jk} &\text{for all } (j,k), j\neq k\\
\sum_{x,\theta}q_{x,\theta}&=1.
\end{align}
\end{subequations}
We note that it is sufficient to optimize over distributions $q_{\vec{x},\vec{\theta}}$ that are symmetric in $\vec{\theta}$: $q_{\vec{x},\vec{\theta}}=q_{\vec{x},-\vec{\theta}}$. This is because, given any decomposition $q_{\vec{x},\vec{\theta}}$ satisfying the constraints, one may always construct the symmetric distribution $\tilde{q}_{\vec{x},\vec{\theta}}\equiv\big(q_{\vec{x},\vec{\theta}}+q_{\vec{x},-\vec{\theta}}\big)/2$ which also satisfies the constraints and produces a lower (or equal) value of the objective: $\mathcal{L}[\tilde{q}_{\vec{x},\vec{\theta}}]\leq\mathcal{L}[q_{\vec{x},\vec{\theta}}]$. Thus, the ultimate goal is to minimize:
\begin{equation}
\mathcal{L}[\tilde{q}_{\vec{x},\vec{\theta}}]=-\sum_{\vec{x},\vec{\theta}}\tilde{q}_{\vec{x},\vec{\theta}}|\bra{\phi_{\vec{x},\vec{\theta}}}\hat{a}\ket{\phi_{\vec{x},\vec{\theta}}}|^2+\left|\left|\langle{\hat{a}^2}\rangle\right|-\sum_{\vec{x},\vec{\theta}}\tilde{q}_{\vec{x},\vec{\theta}}\Re\left(e^{-2i\mu}\bra{\phi_{\vec{x},\vec{\theta}}}\hat{a}\ket{\phi_{\vec{x},\vec{\theta}}}^2\right)\right|,\label{eq:LTildeHighRank}
\end{equation}
subject to the constraints:
\begin{subequations}
\begin{align}
\sum_{\vec{x},\vec{\theta}}\tilde{q}_{\vec{x},\vec{\theta}}x_j^2&=p_j \hspace{1cm} &\text{for all } j\label{eq:HighRankPopConstraint}\\
\sum_{\vec{x},\vec{\theta}}\tilde{q}_{\vec{x},\vec{\theta}}x_jx_k\cos(\theta_j-\theta_k)&=f_{kj}=f_{jk} &\text{for all } (j,k), j\neq k\label{eq:HighRankOffDiagConstraint}\\
\tilde{q}_{\vec{x},\vec{\theta}}&=\tilde{q}_{\vec{x},-\vec{\theta}}\geq0\label{eq:HighRankSymmetricConstraint}\\
\sum_{\vec{x},\vec{\theta}}q_{\vec{x},\vec{\theta}}&=1.\label{eq:HighRankNormalizationConstraint}
\end{align}
\end{subequations}

The absolute in Eq.~\eqref{eq:LTildeHighRank} may multiply what is inside by $\pm 1$, yielding $\mathcal{L}_+[\tilde{q}_{\vec{x},\vec{\theta}}]_+$ or $\mathcal{L}_-[\tilde{q}_{\vec{x},\vec{\theta}}]_-$, each with its own additional inequality constraint:
\begin{subequations}
\begin{align}
\mathcal{L}_+[\tilde{q}_{\vec{x},\vec{\theta}}]=-\sum_{\vec{x},\vec{\theta}}\tilde{q}_{\vec{x},\vec{\theta}}|\bra{\phi_{\vec{x},\vec{\theta}}}\hat{a}\ket{\phi_{\vec{x},\vec{\theta}}}|^2+\left(\left|\langle{\hat{a}^2}\rangle\right|-\sum_{\vec{x},\vec{\theta}}\tilde{q}_{\vec{x},\vec{\theta}}\Re\left(e^{-2i\mu}\bra{\phi_{\vec{x},\vec{\theta}}}\hat{a}\ket{\phi_{\vec{x},\vec{\theta}}}^2\right)\right)\\
\sum_{\vec{x},\vec{\theta}}\tilde{q}_{\vec{x},\vec{\theta}}\Re\left(e^{-2i\mu}\bra{\phi_{\vec{x},\vec{\theta}}}\hat{a}\ket{\phi_{\vec{x},\vec{\theta}}}^2\right)\leq|\langle\hat{a}^2\rangle|,\label{eq:HighRank+Constraint}
\end{align}
\end{subequations}
and
\begin{subequations}
\begin{align}
\mathcal{L}_-[\tilde{q}_{\vec{x},\vec{\theta}}]=-\sum_{\vec{x},\vec{\theta}}\tilde{q}_{\vec{x},\vec{\theta}}|\bra{\phi_{\vec{x},\vec{\theta}}}\hat{a}\ket{\phi_{\vec{x},\vec{\theta}}}|^2-\left(\left|\langle{\hat{a}^2}\rangle\right|-\sum_{\vec{x},\vec{\theta}}\tilde{q}_{\vec{x},\vec{\theta}}\Re\left(e^{-2i\mu}\bra{\phi_{\vec{x},\vec{\theta}}}\hat{a}\ket{\phi_{\vec{x},\vec{\theta}}}^2\right)\right)\\
\sum_{\vec{x},\vec{\theta}}\tilde{q}_{\vec{x},\vec{\theta}}\Re\left(e^{-2i\mu}\bra{\phi_{\vec{x},\vec{\theta}}}\hat{a}\ket{\phi_{\vec{x},\vec{\theta}}}^2\right)\geq|\langle\hat{a}^2\rangle|.\label{eq:HighRank-Constraint}
\end{align}
\end{subequations}
Expressed in this way, the problem resembles a linear programming problem, where $\mathcal{L}_+$ and $\mathcal{L}_-$ are linear objectives, and one optimizes the vector (probability distribution) $\tilde{q}_{\vec{x},\vec{\theta}}$ over the space of possible pure decomposition states $\ket{\phi_{\vec{x},\theta}}$, subject to linear constraints Eq.~\eqref{eq:HighRankPopConstraint}-\ref{eq:HighRankNormalizationConstraint} and Eq.~\eqref{eq:HighRank+Constraint} or Eq.~\eqref{eq:HighRankPopConstraint}-\ref{eq:HighRankNormalizationConstraint} and~\eqref{eq:HighRank-Constraint}. The only caveat is that the space of $\ket{\phi_{\vec{x},\vec{\theta}}}$ is a continuous manifold, and not a discrete set. However, by taking a sufficient discrete sample of this space, we should be able to approximate (tightly upper bound) $\min({\mathcal{L}_+})$ and $\min({\mathcal{L}_-})$. The lesser of these will be an approximation of $\min(\mathcal{L})$. One can also perform iterative methods, improving the decomposition each time by binning around relevant states.

The dimension of the optimized vector will be equivalent to the number of sample states $\ket{\phi_{\vec{x},\vec{\theta}}}$, and thus grows exponentially with $\rank(\hat{\rho})$. In some cases, it is possible to significantly reduce the dimension of the optimized vector by grouping sample states that share parameters (e.g., $\vec{x}$). For example, if $\hat{\rho}$ is a mixture of Fock states (with no photon number coherence), it can be argued that the optimal $\tilde{q}_{\vec{x},\vec{\theta}}$ is a separable distribution in $\vec{x}$ and $\vec{\theta}$, with the optimal $\vec{\theta}$-dependence being known \textit{a priori} (notably, the optimal $\vec{\theta}$-dependence causes the absolute value (``witness") term in $\mathcal{L}$ to vanish, i.e., $\mathcal{L}_+=\mathcal{L}_-$). Then, one needs only to optimize the $\vec{x}$-dependence $\mathcal{Q}(\vec{x})$. Appendix B of Ref. \cite{rogers2024quantifying} describes this in detail.

\section{Effect of Definite Phase Shifts}\label{sec:appendixPhaseShift}
Here we prove that a definite phase shift $e^{-i\hat{n}\chi}$ does not change the ORT measure or metrological power of a mixed state. The simple explanation is that phase shifts merely rotate a state in phase space, whereas the ORT measure and metrological power are ``shape"-dependent quantities.

Let us start with the ORT measure. Since a definite phase shift $e^{-i\hat{n}\chi}$ is a linear optical mapping (and hence a classical operation \cite{ge2020operational}), we have $\mathcal{N}(e^{-i\hat{n}\chi}\hat{\rho}e^{i\hat{n}\chi})\leq\mathcal{N}(\hat{\rho})$, from weak monotonicity (the resource-theoretic property of $\mathcal{N}$). To show that this bound is saturated, let $\{q_j,\ket{\phi_j}\}$ be the optimal decomposition of $\hat{\rho}$, i.e., the decomposition that minimizes the objective on the right hand side of Eq.~\eqref{eq:ORTMixed}. Clearly, $\{q_j,e^{-i\hat{n}\chi}\ket{\phi_j}\}$ is a decomposition of $e^{-i\hat{n}\chi}\hat{\rho}e^{i\hat{n}\chi}$. Since $e^{i\hat{n}\chi}\hat{a}e^{-i\hat{n}\chi}=\hat{a}e^{-i\chi}$,
\begin{equation}
\begin{split}
&\sum_j q_j\bra{\phi_j}e^{i\hat{n}\chi}\hat{a}^\dagger\hat{a}e^{-i\hat{n}\chi}\ket{\phi_j}-\sum_j q_j\left|\bra{\phi_j}e^{i\hat{n}\chi}\hat{a}e^{-i\hat{n}\chi}\ket{\phi_j}\right|^2+\left|\sum_j q_j\bra{\phi_j}e^{i\hat{n}\chi}\hat{a}^2e^{-i\hat{n}\chi}\ket{\phi_j}-\sum_j q_j\bra{\phi_j}e^{i\hat{n}\chi}\hat{a}e^{-i\hat{n}\chi}\ket{\phi_j}^2\right|\\
&=\sum_j q_j\bra{\phi_j}\hat{a}^\dagger\hat{a}\ket{\phi_j}-\sum_j q_j\left|\bra{\phi_j}\hat{a}\ket{\phi_j}\right|^2+\left|\sum_j q_j\bra{\phi_j}\hat{a}^2\ket{\phi_j}-\sum_j q_j\bra{\phi_j}\hat{a}\ket{\phi_j}^2\right|\\
&=\mathcal{N}(\hat{\rho}).
\end{split}
\end{equation}
Since $\{q_j,e^{-i\hat{n}\chi}\ket{\phi_j}\}$ achieves the upper bound value set by $\mathcal{N}(\hat{\rho})$, it must be optimal, and:
\begin{equation}
\mathcal{N}(e^{-i\hat{n}\chi}\hat{\rho}e^{i\hat{n}\chi})=\mathcal{N}(\hat{\rho}).
\end{equation}

Next we consider the metrological power (of quadrature variance), defined as $\mathcal{M}(\hat{\rho})\equiv\max[F_X(\hat{\rho})-1/2,0]$, where
\begin{equation}
F_X(\hat{\rho})=\max_\mu\left[\min_{\{q_j,\ket{\phi_j}\}}\left(\sum_jq_j\bra{\phi_j}(\Delta \hat{X}_\mu)^2\ket{\phi_j}\right)\right]
\end{equation}
is the Quantum Fisher Information for the optimal quadrature. Here $\hat{X}_\mu=i(e^{-i\mu}\hat{a}^ \dagger-e^{i\mu}\hat{a})/\sqrt{2}$. From $e^{i\hat{n}\chi}\hat{a}e^{-i\hat{n}\chi}=\hat{a}e^{-i\chi}$, it follows that $e^{i\hat{n}\chi}\hat{X}_\mu e^{-i\hat{n}\chi}=\hat{X}_{\mu-\chi}$. Thus, a definite phase shift creates a one-to-one mapping between quadratures. Clearly, a definite phase shift also creates a one-to-one mapping between decompositions $\{q_j,\ket{\phi_j}\}$ of $\hat{\rho}$ and decompositions $\{q_j,e^{-i\hat{n}\chi}\ket{\phi_j}\}$ of $e^{-i\hat{n}\chi}\hat{\rho}e^{i\hat{n}\chi}$. Combining these, we may define $\Theta_\chi((\hat{X}_\mu,\{q_j,\ket{\phi_j}\}))=(\hat{X}_{\mu+\chi},\{q_j,e^ {-i\hat{n}\chi}\ket{\phi_j}\})$ as a one-to-one mapping between quadrature-decomposition pairs of $\hat{\rho}$ and quadrature-decomposition pairs of $e^{-i\hat{n}\chi}\hat{\rho}e^{i\hat{n}\chi}$. We may also define the scalar objective function: $\mathcal{L}((\hat{X}_\mu,\{q_j,\ket{\phi_j}\}))=\sum_jq_j\bra{\phi_j}(\Delta \hat{X}_\mu)^2\ket{\phi_j}$.

Since:
\begin{equation}
\begin{split}
\sum_jq_j\bra{\phi_j}e^{i\hat{n}\chi}(\Delta \hat{X}_{\mu+\chi})^2e^{-i\hat{n}\chi}\ket{\phi_j}&=\sum_jq_j\bra{\phi_j}e^{i\hat{n}\chi}\hat{X}_{\mu+\chi}^2e^{-i\hat{n}\chi}\ket{\phi_j}-\sum_jq_j\bra{\phi_j}e^{i\hat{n}\chi}\hat{X}_{\mu+\chi}e^{-i\hat{n}\chi}\ket{\phi_j}^2\\
&=\sum_jq_j\bra{\phi_j}\hat{X}_{\mu}^2\ket{\phi_j}-\sum_jq_j\bra{\phi_j}\hat{X}_{\mu}\ket{\phi_j}^2\\
&=\sum_jq_j\bra{\phi_j}(\Delta \hat{X}_\mu)^2\ket{\phi_j},
\end{split}
\end{equation}
we have $\mathcal{L}(\Theta_\chi((\hat{X}_\mu,\{q_j,\ket{\phi_j}\})))=\mathcal{L}((\hat{X}_\mu,\{q_j,\ket{\phi_j}\})).$ If the optimal quadrature-decomposition pair of $\hat{\rho}$ is $(\hat{X}_\mu,\{q_j,\ket{\phi_j}\})$, then the optimal quadrature-decomposition pair of $e^{-i\hat{n}\chi}\hat{\rho}e^{i\hat{n}\chi}$ is $\Theta_\chi((\hat{X}_\mu,\{q_j,\ket{\phi_j}\}))$, and $\mathcal{M}(\hat{\rho})=\mathcal{M}(e^{-i\hat{n}\chi}\hat{\rho}e^{i\hat{n}\chi})$.

\section{Photon Loss Channel}\label{sec:appendixPhotonLossChannel}

Here we show that Assumptions 1 and 2 can describe a pure state subject to weak photon loss $\theta^2\langle\hat{a}^\dagger\hat{a}\rangle\ll1$, where $\theta^2$ is the reflectivity). We also provide expressions for the associated ORT measure to order $\theta^2$. We start with a definite parity pure state with all real and positive photon number amplitudes: $\ket{\psi_0}=\sum_{n=0}^\infty\ket{2n+b}\braket{2n+b|\psi_0}$, where $b\in\{0,1\}$ and $\braket{2n+b|\psi_0}\geq0$ for all $n$ (we could also allow for a definite phase shift $e^{i\hat{n}\phi}$, but do not for simplicity). We furthermore assume $\ket{\psi_0}\neq\ket{0}$. We model photon loss by combining $\ket{\psi_0}$ with vacuum at a beamsplitter with $\hat{U}(\theta)=\exp\left(\theta(-\hat{a}^\dagger\hat{b}+\hat{a}\hat{b}^\dagger)\right)$. Up to the second order in $\theta$, we have 
\begin{equation}
\begin{split}
\hat{U}(\theta)\ket{\psi_0}\ket{0}&=\left[1+\theta\left(-\hat{a}^\dagger\hat{b}+\hat{a}\hat{b}^\dagger\right)+\frac{\theta^2}{2}\left(-\hat{a}^\dagger\hat{b}+\hat{a}\hat{b}^\dagger\right)^2+\mathcal{O}(\theta^3)\right]\ket{\psi_0}\ket{0}\\
&=\ket{\psi_0}\ket{0}+\theta\hat{a}\ket{\psi_0}\ket{1}+\frac{\theta^2}{2}\left(\sqrt{2}\hat{a}^2\ket{\psi_0}\ket{2}-\hat{a}^\dagger\hat{a}\ket{\psi_0}\ket{0}\right)+\mathcal{O}(\theta^3).
\end{split}
\end{equation}
So
\begin{equation}
\begin{split}
&\hat{U}(\theta)\ket{\psi_0}\bra{\psi_0}\otimes\ket{0}\bra{0}\hat{U}^\dagger(\theta)\\
&=\ket{\psi_0}\bra{\psi_0}\otimes\ket{0}\bra{0}+\theta\left(\ket{\psi_0}\bra{\psi_0}\hat{a}^\dagger\otimes\ket{0}\bra{1}+\hat{a}\ket{\psi_0}\bra{\psi_0}\otimes\ket{1}\bra{0}\right)\\
&+\frac{\theta^2}{2}\left[\left(-\ket{\psi_0}\bra{\psi_0}\hat{a}^\dagger\hat{a}-\hat{a}^\dagger\hat{a}\ket{\psi_0}\bra{\psi_0}\right)\otimes\ket{0}\bra{0}+2\hat{a}\ket{\psi_0}\bra{\psi_0}\hat{a}^\dagger\otimes\ket{1}\bra{1}+\sqrt{2}\ket{\psi_0}\bra{\psi_0}\hat{a}^{\dagger2}\otimes\ket{0}\bra{2}+\sqrt{2}\hat{a}^2\ket{\psi_0}\bra{\psi_0}\otimes\ket{2}\bra{0}\right]\\
&+\mathcal{O}(\theta^3).
\end{split}
\end{equation}
The reduced state of the transmitted mode will be:
\begin{equation}
\begin{split}
\hat{\rho}(\theta)&=\Tr_B\left[\hat{U}(\theta)\ket{\psi_0}\bra{\psi_0}\otimes\ket{0}\bra{0}\hat{U}^\dagger(\theta)\right]\\
&=\ket{\psi_0}\bra{\psi_0}+\theta^2\left[\left(-\ket{\psi_0}\bra{\psi_0}\frac{\hat{a}^\dagger\hat{a}}{2}-\frac{\hat{a}^\dagger\hat{a}}{2}\ket{\psi_0}\bra{\psi_0}\right)+\hat{a}\ket{\psi_0}\bra{\psi_0}\hat{a}^\dagger\right]+\mathcal{O}(\theta^3)\\
&=\left(1-\frac{\theta^2\hat{a}^\dagger\hat{a}}{2}\right)\ket{\psi_0}\bra{\psi_0}\left(1-\frac{\theta^2\hat{a}^\dagger\hat{a}}{2}\right)+\theta^2\hat{a}\ket{\psi_0}\bra{\psi_0}\hat{a}^\dagger+\mathcal{O}(\theta^3).
\end{split}
\end{equation}
The $\theta^2$ term in the second line is recognizable as a Linblandian update $\theta^2\mathcal{L}\left(\ket{\psi_0}\bra{\psi_0}\right)$ \cite{wiseman2009quantum}. One may question the introduction of the $\frac{\theta^4}{4}\hat{a}^\dagger\hat{a}\ket{\psi_0}\bra{\psi_0}\hat{a}^\dagger\hat{a}$ term in the last line. We do not claim this is the only $\theta^4$ term in the full expansion --- the ``catch-all" $\mathcal{O}(\theta^3)$ may include more. Since $\ket{\psi_0}$ has definite parity, $\left(1-\frac{\theta^2\hat{a}^\dagger\hat{a}}{2}\right)\ket{\psi_0}$ and $\hat{a}\ket{\psi_0}$ are orthogonal. This suggests a rank-two approximation $\hat{\rho}_\text{approx}(\theta)\approx\hat{\rho}(\theta)$ in the $\theta^2\bra{\psi_0}\hat{a}^\dagger\hat{a}\ket{\psi_0}\ll1$ regime, where:
\begin{align}
\hat{\rho}_\text{approx}(\theta)&=\left(1-\theta^2\bra{\psi_0}\hat{a}^\dagger\hat{a}\ket{\psi_0}\right)\ket{\psi_1}\bra{\psi_1}+\theta^2\bra{\psi_0}\hat{a}^\dagger\hat{a}\ket{\psi_0}\ket{\psi_2}\bra{\psi_2}\label{eq:rhoApprox1}\\
\hat{\rho}_\text{approx}(\theta)&=\hat{\rho}(\theta)+\mathcal{O}(\theta^3).\label{eq:rhoApprox2}
\end{align}
$\ket{\psi_1}$ and $\ket{\psi_2}$ are orthonormal eigenvectors of $\hat{\rho}_\text{approx}$. Note that $\hat{\rho}_\text{approx}(\theta)$, $\ket{\psi_1}$ and $\ket{\psi_2}$ are not determined uniquely by Eq. \eqref{eq:rhoApprox1}--\eqref{eq:rhoApprox2}. Eq. \eqref{eq:rhoApprox2} only requires that $\hat{\rho}_\text{approx}(\theta)$ and $\hat{\rho}(\theta)$ be the same to order $\theta^2$, providing flexibility at higher orders. That being said, a natural choice is:
\begin{align}
\ket{\psi_1}&=\frac{\left(1-\frac{\theta^2\hat{a}^\dagger\hat{a}}{2}\right)\ket{\psi_0}}{\sqrt{\bra{\psi_0}\left(1-\frac{\theta^2\hat{a}^\dagger\hat{a}}{2}\right)^2\ket{\psi_0}}}\\
\ket{\psi_2}&=\frac{\hat{a}\ket{\psi_0}}{\sqrt{\bra{\psi_0}\hat{a}^\dagger\hat{a}\ket{\psi_0}}}.
\end{align}
An alternative choice (one perhaps more in line with the examples in the main text) would be to use instead $\ket{\psi_2'}\propto\hat{a}\ket{\psi_1}$. Let us now ask whether $\ket{\psi_1}$ and $\ket{\psi_2}$ satisfy Assumptions 1 and 2 in the main text, for sufficiently small (but nonzero) $\theta$. Since $\ket{\psi_0}$ has definite parity, $\ket{\psi_1}$ and $\ket{\psi_2}$ have definite parity such that $\bra{\psi_1}\hat{a}\ket{\psi_1}=\bra{\psi_2}\hat{a}\ket{\psi_2}=0$ and Assumption 1 is satisfied. To check Assumption 2, we recall that $\braket{n|\psi_0}\geq0$ for all $n$, $\ket{\psi_0}\neq\ket{0}$, and make a table.

\begin{center}
\begin{tabular}{ c|c|c|c|c } 
  & $\bra{\psi_1}\hat{a}^2\ket{\psi_1}$ & $\bra{\psi_2}\hat{a}^2\ket{\psi_2}$ & $\bra{\psi_1}\hat{a}\ket{\psi_2}$ & $\bra{\psi_2}\hat{a}\ket{\psi_1}$
 \\
 \hline
 Proportional to& $\bra{\psi_0}\left(1-\frac{\theta^2\hat{a}^\dagger\hat{a}}{2}\right)\hat{a}^2\left(1-\frac{\theta^2\hat{a}^\dagger\hat{a}}{2}\right)\ket{\psi_0}$ & $\bra{\psi_0}\hat{a}^\dagger\hat{a}^3\ket{\psi_0}$ & $\bra{\psi_0}\left(1-\frac{\theta^2\hat{a}^\dagger\hat{a}}{2}\right)\hat{a}^2\ket{\psi_0}$ & $\bra{\psi_0}\hat{a}^\dagger\hat{a}\left(1-\frac{\theta^2\hat{a}^\dagger\hat{a}}{2}\right)\ket{\psi_0}$ \\
 \hline
 If $\bra{\psi_0}\hat{a}^2\ket{\psi_0}>0$ & $> 0$ (sufficiently small $\theta$) & $\geq0$ & $> 0$ (suff. small $\theta$) & $> 0$ (suff. small $\theta$)\\ 
 \hline
 If $\bra{\psi_0}\hat{a}^2\ket{\psi_0}=0$ & $0$ & $0$ & $0$ & $> 0$ (suff. small $\theta$)
\end{tabular}
\end{center}
For example, the second column, third row cell should be interpreted as: if $\bra{\psi_0}\hat{a}^2\ket{\psi_0}>0$, there exists $\theta_0$ such that for all positive $\theta<\theta_0$, $\bra{\psi_1}\hat{a}^2\ket{\psi_1}>0$. One can check this table against Assumption 2 to confirm $\ket{\psi_1}$ and $\ket{\psi_2}$ satisfy Assumption 2 for sufficiently small $\theta$.

Furthermore, we describe the behavior the ORT measure of nonclassicality under weak photon loss. Recall that $\hat{\rho}(\theta)$ and $\hat{\rho}_\text{approx}(\theta)$ are equivalent to order $\theta^2$ (Eq. \eqref{eq:rhoApprox2}). To calculate $\mathcal{N}\left(\hat{\rho}(\theta)\right)$ to order $\theta^2$, it should then suffice to calculate $\mathcal{N}\left(\hat{\rho}_\text{approx}\right)$ to order $\theta^2$ (assuming $\mathcal{N}(\hat{\rho}(\theta))$ is well-behaved around $\theta=0$):
\begin{equation}
\mathcal{N}(\hat{\rho}(\theta))=\mathcal{N(\hat{\rho}_\text{approx}(\theta))}+\mathcal{O}(\theta^3).
\end{equation}
Since the eigenstates $\ket{\psi_1}$ and $\ket{\psi_2}$ satisfy Assumptions 1 and 2 for sufficiently small $\theta$, we may use Eq. \eqref{eq:overSqueezed}--\eqref{eq:underSqueezed}.
The necessary quantities are:
\begin{align}
\bra{\psi_1}\hat{a}^\dagger\hat{a}\ket{\psi_1}&=\frac{\bra{\psi_0}\left(1-\frac{\theta^2\hat{a}^\dagger\hat{a}}{2}\right)\hat{a}^\dagger\hat{a}\left(1-\frac{\theta^2\hat{a}^\dagger\hat{a}}{2}\right)\ket{\psi_0}}{\bra{\psi_0}\left(1-\frac{\theta^2\hat{a}^\dagger\hat{a}}{2}\right)^2\ket{\psi_0}}=\left(\bra{\psi_0}\hat{a}^\dagger\hat{a}-\theta^2\hat{a}^\dagger\hat{a}\hat{a}^\dagger\hat{a}\ket{\psi_0}\right)\left(1+\theta^2\bra{\psi_0}\hat{a}^\dagger\hat{a}\ket{\psi_0}\right)+\mathcal{O}(\theta^3)\\
\bra{\psi_2}\hat{a}^\dagger\hat{a}\ket{\psi_2}&=\frac{\bra{\psi_0}\hat{a}^\dagger\hat{a}^\dagger\hat{a}\hat{a}\ket{\psi_0}}{\bra{\psi_0}\hat{a}^\dagger\hat{a}\ket{\psi_0}}\\
\bra{\psi_1}\hat{a}^2\ket{\psi_1}&=\frac{\bra{\psi_0}\left(1-\frac{\theta^2\hat{a}^\dagger\hat{a}}{2}\right)\hat{a}^2\left(1-\frac{\theta^2\hat{a}^\dagger\hat{a}}{2}\right)\ket{\psi_0}}{\bra{\psi_0}\left(1-\frac{\theta^2\hat{a}^\dagger\hat{a}}{2}\right)^2\ket{\psi_0}}=\left(\bra{\psi_0}\hat{a}^2-\frac{\theta^2}{2}\{\hat{a}^\dagger\hat{a},\hat{a}^2\}\ket{\psi_0}\right)\left(1+\theta^2\bra{\psi_0}\hat{a}^\dagger\hat{a}\ket{\psi_0}\right)+\mathcal{O}(\theta^3)\\
\bra{\psi_2}\hat{a}^2\ket{\psi_2}&=\frac{\bra{\psi_0}\hat{a}^\dagger\hat{a}^3\ket{\psi_0}}{\bra{\psi_0}\hat{a}^\dagger\hat{a}\ket{\psi_0}}\\
\bra{\psi_1}\hat{a}\ket{\psi_2}&=\frac{\bra{\psi_0}\left(1-\frac{\theta^2\hat{a}^\dagger\hat{a}}{2}\right)\hat{a}^2\ket{\psi_0}}{\sqrt{\bra{\psi_0}\left(1-\frac{\theta^2\hat{a}^\dagger\hat{a}}{2}\right)^2\ket{\psi_0}\bra{\psi_0}\hat{a}^\dagger\hat{a}\ket{\psi_0}}}=\left(\bra{\psi_0}\hat{a}^2-\frac{\theta^2}{2}\hat{a}^\dagger\hat{a}^3\ket{\psi_0}\right)\left(\frac{1+\frac{\theta^2}{2}\bra{\psi_0}\hat{a}^\dagger\hat{a}\ket{\psi_0}}{\sqrt{\bra{\psi_0}\hat{a}^\dagger\hat{a}\ket{\psi_0}}}\right)+\mathcal{O}(\theta^3)\\
\bra{\psi_2}\hat{a}\ket{\psi_1}&=\frac{\bra{\psi_0}\hat{a}^\dagger\hat{a}\left(1-\frac{\theta^2\hat{a}^\dagger\hat{a}}{2}\right)\ket{\psi_0}}{\sqrt{\bra{\psi_0}\left(1-\frac{\theta^2\hat{a}^\dagger\hat{a}}{2}\right)^2\ket{\psi_0}\bra{\psi_0}\hat{a}^\dagger\hat{a}\ket{\psi_0}}}=\left(\bra{\psi_0}\hat{a}^\dagger\hat{a}-\frac{\theta^2}{2}\hat{a}^\dagger\hat{a}\hat{a}^\dagger\hat{a}\ket{\psi_0}\right)\left(\frac{1+\frac{\theta^2}{2}\bra{\psi_0}\hat{a}^\dagger\hat{a}\ket{\psi_0}}{\sqrt{\bra{\psi_0}\hat{a}^\dagger\hat{a}\ket{\psi_0}}}\right)+\mathcal{O}(\theta^3)
\end{align}
First, let us suppose that $\bra{\psi_0}\hat{a}^2\ket{\psi_0}>0$. Then, for sufficiently small $\theta$, $\langle\hat{a}^2\rangle=\bra{\psi_0}\hat{a}^2\ket{\psi_0}+\mathcal{O}(\theta^2)>\theta^2\bra{\psi_0}\hat{a}^\dagger\hat{a}\ket{\psi_0}(1-\theta^2\bra{\psi_0}\hat{a}^\dagger\hat{a}\ket{\psi_0})\left(\bra{\psi_2}\hat{a}\ket{\psi_1}+\bra{\psi_1}\hat{a}\ket{\psi_2}\right)^2$ and we may use Eq. \eqref{eq:overSqueezed} for the ORT measure.
\begin{equation}
\begin{split}
\mathcal{N}\left(\hat{\rho}_\text{approx}(\theta)\right)&=\left(1-\theta^2\bra{\psi_0}\hat{a}^\dagger\hat{a}\ket{\psi_0}\right)\left(\bra{\psi_1}\hat{a}^\dagger\hat{a}+\hat{a}^2\ket{\psi_1}\right)+\theta^2\bra{\psi_0}\hat{a}^\dagger\hat{a}\ket{\psi_0}\left(\bra{\psi_2}\hat{a}^\dagger\hat{a}+\hat{a}^2\ket{\psi_2}\right)\\
&-2\theta^2\bra{\psi_0}\hat{a}^\dagger\hat{a}\ket{\psi_0}\left(1-\theta^2\bra{\psi_0}\hat{a}^\dagger\hat{a}\ket{\psi_0}\right)\left(\bra{\psi_1}\hat{a}\ket{\psi_2}+\bra{\psi_2}\hat{a}\ket{\psi_1}\right)^2+\mathcal{O}(\theta^3)\\
&=\bra{\psi_0}\hat{a}^\dagger\hat{a}+\hat{a}^2-\theta^2\left(\hat{a}^\dagger\hat{a}\hat{a}^\dagger\hat{a}+\frac{\{\hat{a}^\dagger\hat{a},\hat{a}^2\}}{2}\right)\ket{\psi_0}+\theta^2\bra{\psi_0}\hat{a}^\dagger\hat{a}^\dagger\hat{a}\hat{a}+\hat{a}^\dagger\hat{a}^3\ket{\psi_0}\\
&-2\theta^2\bra{\psi_0}\hat{a}^\dagger\hat{a}+\hat{a}^2\ket{\psi_0}^2+\mathcal{O}(\theta^3)\\
&=\bra{\psi_0}\hat{a}^\dagger\hat{a}+\hat{a}^2\ket{\psi_0}-\theta^2\left(\bra{\psi_0}\hat{a}^\dagger\hat{a}+\hat{a}^2\ket{\psi_0}+2\bra{\psi_0}\hat{a}^\dagger\hat{a}+\hat{a}^2\ket{\psi_0}^2\right)+\mathcal{O}(\theta^3).
\end{split}
\end{equation}
At last we have:
\begin{equation}
\mathcal{N}\left(\hat{\rho}(\theta)\right)=\mathcal{N}\left(\hat{\rho}(0)\right)-\theta^2\left(\mathcal{N}\left(\hat{\rho}(0)\right)+2\mathcal{N}\left(\hat{\rho}(0)\right)^2\right)+\mathcal{O}(\theta^3), \hspace{0.5in} \text{if $\bra{\psi_0}\hat{a}^2\ket{\psi_0}>0$.}\label{eq:secondOrderApproximation1}
\end{equation}
Since we used Eq. ~\eqref{eq:overSqueezed}, the metrological power $\mathcal{M}$ will be the same to order $\theta^2$ (Eq. \eqref{eq:M1}). Next let us suppose that $\bra{\psi_0}\hat{a}^2\ket{\psi_0}=0$. Then $\langle\hat{a}^2\rangle=\mathcal{O}(\theta^3)$, $\bra{\psi_1}\hat{a}\ket{\psi_2}=\mathcal{O}(\theta^2)$, $\bra{\psi_2}\hat{a}\ket{\psi_1}=\sqrt{\bra{\psi_0}\hat{a}^\dagger\hat{a}\ket{\psi_0}}+\mathcal{O}(\theta^2)$. For sufficiently small $\theta$, $\langle\hat{a}^2\rangle<\theta^2\bra{\psi_0}\hat{a}^\dagger\hat{a}\ket{\psi_0}(1-\theta^2\bra{\psi_0}\hat{a}^\dagger\hat{a}\ket{\psi_0})\left(\bra{\psi_2}\hat{a}\ket{\psi_1}+\bra{\psi_1}\hat{a}\ket{\psi_2}\right)^2=\theta^2\bra{\psi_0}\hat{a}^\dagger\hat{a}\ket{\psi_0}^2+\mathcal{O}(\theta^3)$ and we may use Eq. \eqref{eq:underSqueezed} for the ORT measure. We obtain:
\begin{equation}
\begin{split}
\mathcal{N}\left(\hat{\rho}_\text{approx}(\theta)\right)&=\left(1-\theta^2\bra{\psi_0}\hat{a}^\dagger\hat{a}\ket{\psi_0}\right)\bra{\psi_1}\hat{a}^\dagger\hat{a}\ket{\psi_1}+\theta^2\bra{\psi_0}\hat{a}^\dagger\hat{a}\ket{\psi_0}\bra{\psi_2}\hat{a}^\dagger\hat{a}\ket{\psi_2}\\
&-\theta^2\bra{\psi_0}\hat{a}^\dagger\hat{a}\ket{\psi_0}(1-\theta^2\bra{\psi_0}\hat{a}^\dagger\hat{a}\ket{\psi_0})\frac{\left(\bra{\psi_2}\hat{a}\ket{\psi_1}^2-\bra{\psi_1}\hat{a}\ket{\psi_2}^2\right)^2}{\bra{\psi_2}\hat{a}\ket{\psi_1}^2+\bra{\psi_1}\hat{a}\ket{\psi_2}^2}+\mathcal{O}(\theta^3)\\
&=\bra{\psi_0}\hat{a}^\dagger\hat{a}\ket{\psi_0}-\theta^2\left(\bra{\psi_0}\hat{a}^\dagger\hat{a}\ket{\psi_0}+\bra{\psi_0}\hat{a}^\dagger\hat{a}\ket{\psi_0}^2\right)+\mathcal{O}(\theta^3)
\end{split}
\end{equation}
So we have:
\begin{equation}
\mathcal{N}\left(\hat{\rho}(\theta)\right)=\mathcal{N}\left(\hat{\rho}(0)\right)-\theta^2\left(\mathcal{N}\left(\hat{\rho}(0)\right)+\mathcal{N}\left(\hat{\rho}(0)\right)^2\right)+\mathcal{O}(\theta^3), \hspace{0.5in} \text{if $\bra{\psi_0}\hat{a}^2\ket{\psi_0}=0$.}\label{eq:secondOrderApproximation2}
\end{equation}
Since we used Eq. ~\eqref{eq:underSqueezed}, the metrological power $\mathcal{M}$ will generally be different at order $\theta^2$ (Eq. \eqref{eq:M2}). Both of these scenarios show the trend of initial decreasing nonclassicality as it becomes mixed due to weak photon losses (see Figs.~(\ref{fig:cats}-\ref{fig:ORT_indef_parity}) in the main text). 

\section{Additional Examples}
\label{sec:appendixAdditionalExamples}
Here we include additional examples, which did not make it into the final version of the main text.

\subsection{Rank-2 Partially-coherent Mixtures of Three Fock States}
Partially-coherent mixtures of three neighboring Fock states of the form:
\begin{equation}
\hat{\rho}=
\begin{pmatrix}
    p_{n+2} & f\sqrt{p_{n+2}p_{n+1}} & \sqrt{p_{n+2}p_n}\\
    f\sqrt{p_{n+2}p_{n+1}} & p_{n+1} & f\sqrt{p_{n+1}p_n}\\
    \sqrt{p_{n+2}p_n} & f\sqrt{p_{n+1}p_n} & p_n
\end{pmatrix}
\label{eq:partiallyCoherentMixture3FockStates}
\end{equation}
are rank-two (for $f<1$) and satisfy Assumptions 1*, 2*, 3, and 4. The special basis states $\ket{\psi_2}$ and $\ket{\psi_1}$ are:
\begin{subequations}
\begin{align}
\ket{\psi_2}&\equiv\sqrt{\frac{p_{n+2}}{p_{n+2}+p_n}}\ket{n+2}+\sqrt{\frac{p_n}{p_{n+2}+p_n}}\ket{n}\\
\ket{\psi_1}&\equiv\ket{n+1},
\end{align}
\end{subequations}
with populations $\tilde{p}_2=p_{n+2}+p_n$ and $\tilde{p}_1=p_{n+1}$, and coherence strength $f$ (see Eq.~\eqref{eq:partiallyCoherentMixture}). For simplicity, we take $f$, between $0$ and $1$, to be real. However, one could apply a phase shift $e^{i\hat{n}\chi}$ to $\hat{\rho}$ to make the off-diagonal elements in Eq.~\eqref{eq:partiallyCoherentMixture3FockStates} complex, albeit with constraints. Such a phase shift would not change the ORT measure.

One finds that:
\begin{subequations}
\begin{align}
\bra{\psi_1}\hat{a}\ket{\psi_2}&=\sqrt{\frac{p_{n+2}(n+2)}{p_{n+2}+p_n}}\\
\bra{\psi_2}\hat{a}\ket{\psi_1}&=\sqrt{\frac{p_{n}(n+1)}{p_{n+2}+p_n}}\\
\bra{\psi_1}\hat{a}^2\ket{\psi_1}&=0\\
\bra{\psi_2}\hat{a}^2\ket{\psi_2}&=\sqrt{\frac{p_{n+2}p_n(n+2)(n+1)}{(p_{n+2}+p_{n})^2}}\\
\bra{\psi_1}\hat{a}^2\ket{\psi_2}&=0\\
\bra{\psi_2}\hat{a}^2\ket{\psi_1}&=0.
\end{align}
\end{subequations}

\begin{figure}
    \centering
     \captionsetup{justification=raggedright, singlelinecheck=false}
    \includegraphics[width=\columnwidth]{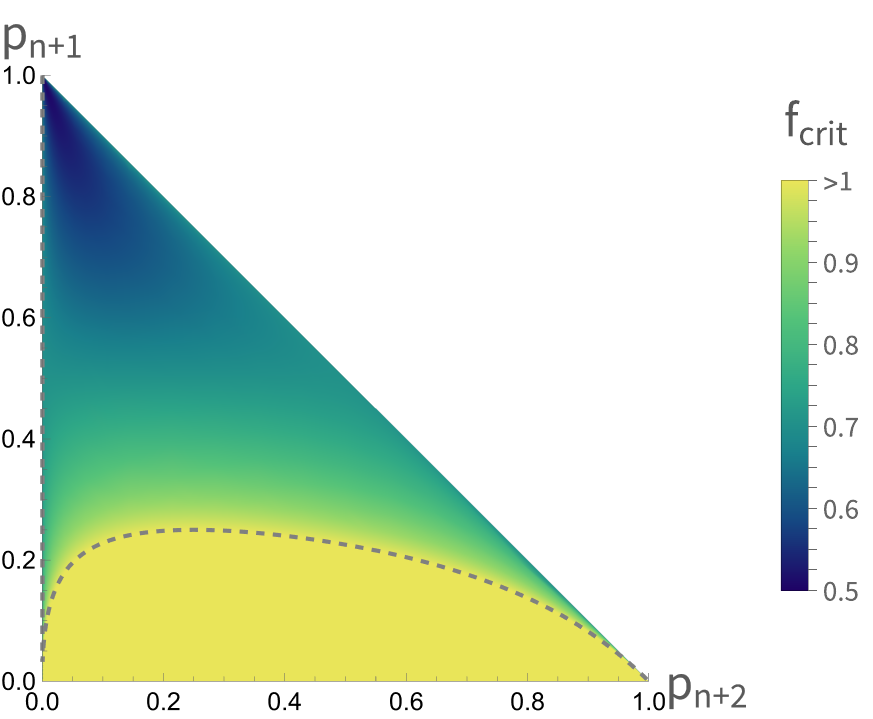}
    \caption{This figure shows which nonclassicality equation holds , among Eq.~\eqref{eq:overSqueezed2}--\eqref{eq:underSqueezedB}, for the rank-2 partially-coherent mixture of three neighboring Fock states parameterized in Eq.~\eqref{eq:partiallyCoherentMixture3FockStates}. $p_n$ is fixed by normalization and the figure assumes $n=0$. Here, $f_\text{crit}$ is the positive value satisfying $|\langle\hat{a}^2\rangle|=\eta(f_\text{crit},p)$. For populations $(p_{n+2},p_{n+1})$ below the dotted line, $f_\text{crit}>1$; since $f$ can never take such a value, Eq.~\eqref{eq:overSqueezed2} always holds. Above the dotted line, Eq.~\eqref{eq:underSqueezedB} holds for $f\geq f_\text{crit}$ and Eq.~\eqref{eq:underSqueezedA} holds for $f<f_\text{crit}$.}
    \label{fig:rank2within3Regions}
\end{figure}

Fig. \ref{fig:rank2within3Regions} shows which nonclassicality equation, among Eq.~\eqref{eq:overSqueezed2}--\eqref{eq:underSqueezedB}, holds, for each choice of photon number populations and coherence. Here, $f_\text{crit}$ is the positive value satisfying $|\langle\hat{a}^2\rangle|=\eta(f_\text{crit},p)$. For populations $(p_{n+2},p_{n+1})$ below the dotted line, $f_\text{crit}>1$; since $f$ can never take such a value, Eq.~\eqref{eq:overSqueezed2} always holds. Above the dotted line, Eq.~\eqref{eq:underSqueezedB} holds for $f\geq f_\text{crit}$ and Eq.~\eqref{eq:underSqueezedA} holds for $f<f_\text{crit}$.

\subsection{Mixture of Three Cat states}
Lastly, we consider a rank-three mixture of three cat states (see Appendix \ref{sec:appendixHighRank} for methodology):
\begin{equation}
\hat{\rho}=p_0\ket{\text{cat}_0}\bra{\text{cat}_0}+p_1\ket{\text{cat}}\bra{\text{cat}_1}+p_2\ket{\text{cat}_2}\bra{\text{cat}_2},\label{eq:threeCatMixture}
\end{equation}
where:
\begin{subequations}
\begin{align}
\ket{\text{cat}_0}&=\frac{1}{\sqrt{N_0}}\left(\ket{\alpha}+\ket{e^{2\pi i/3}\alpha}+\ket{e^{4\pi i/3}\alpha}\right)\label{eq:cat0of3}\\
\ket{\text{cat}_1}&=\frac{1}{\sqrt{N_1}}\left(\ket{\alpha}+e^{2\pi i/3}\ket{e^{2\pi i/3}\alpha}+e^{4\pi i/3}\ket{e^{4\pi i/3}\alpha}\right)\label{eq:cat1of3}\\
\ket{\text{cat}_2}&=\frac{1}{\sqrt{N_2}}\left(\ket{\alpha}+e^{4\pi i/3}\ket{e^{2\pi i/3}\alpha}+e^{2\pi i/3}\ket{e^{4\pi i/3}\alpha}\right),\label{eq:cat2of3}
\end{align}
\end{subequations}
$\ket{\alpha}$ is a coherent state, and the $N_i$ are normalization factors. These cat states are mutually orthogonal, and cycled by $\hat{a}$: $\hat{a}\ket{\text{cat}}_i\propto\ket{\text{cat}}_{i+1\mod3}$. Also, each consists only of photon numbers separated by 3: $\braket{n|\text{cat}_i}\propto\delta_{0,n+i\mod3}$.

\begin{figure}
    \centering
     \captionsetup{justification=raggedright, singlelinecheck=false}
    \includegraphics[width=\columnwidth]{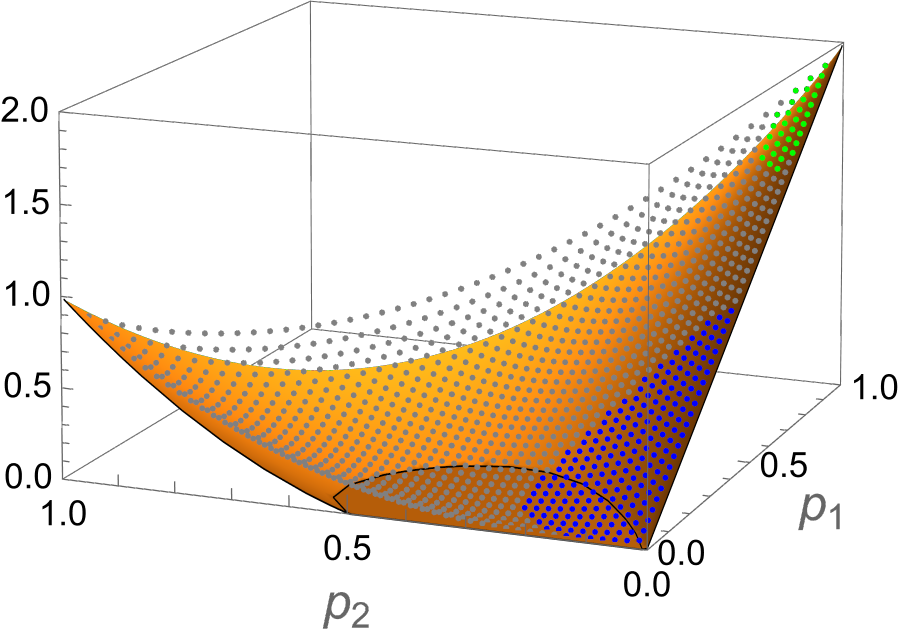}
    \caption{ORT measure (dots) and metrological power (orange surface) of a mixture of three cat states (Eq.~\eqref{eq:threeCatMixture}--\eqref{eq:cat2of3}, with $\alpha=0.5$), as a function of populations. The ORT measure's three regions, represented by different colors, are distinguished by their optimal decompositions, as discussed in the main text. To clarify, the ORT measure is (slightly) 
    greater than zero when $p_1=p_2=0.$ The metrological power is zero in the region outlined by a black curve.}
    \label{fig:threeMixedCats}
\end{figure}

The ORT measure $\mathcal{N}$ and the metrological power $\mathcal{M}$ for arbitrary populations, and $\alpha=0.5$, are plotted in Fig. \ref{fig:threeMixedCats}.

The ORT measure results are divided into three regions. In the largest region (gray), the optimal decomposition uses three states: \begin{equation}
\ket{\phi_{T,j}}=\sqrt{p_0}\ket{\text{cat}_0}+\sqrt{p_1}\mu^j\ket{\text{cat}_1}+\sqrt{p_2}\mu^{2j}\ket{\text{cat}_2},    
\end{equation}
where $j\in\{0,1,2\}$ and $\mu=e^{2\pi i/3}$, with equal probabilities $1/3$ each. In the second largest region (blue), the decomposition uses three states of the form:
\begin{equation}
\begin{split}
\ket{\phi_{U,j}}&=\sqrt{1-q}\ket{\text{cat}_0}\\
&+\sqrt{q}\left(\sqrt{\frac{p_1}{p_1+p_2}}\mu^j\ket{\text{cat}_1}+\sqrt{\frac{p_2}{p_1+p_2}}\mu^{2j}\ket{\text{cat}_2}\right),
\end{split}
\end{equation}
where $j\in\{0,1,2\}$, with equal probabilities $\frac{p_1+p_2}{3q}$ each, and the state $\ket{\text{cat}_0}$, with probability $(1-p_1-p_2)$. The particular value of $q$ is determined by optimization. Lastly, in the smallest region (green), the decomposition uses three states of the form:
\begin{equation}
\begin{split}
\ket{\phi_{L,j}}&=\sqrt{q'}\left(\sqrt{\frac{p_0}{p_0+p_2}}\ket{\text{cat}_0}+\sqrt{\frac{p_2}{p_0+p_2}}\mu^j\ket{\text{cat}_2}\right)\\
&+\sqrt{1-q'}\mu^{2j}\ket{\text{cat}_1},
\end{split}
\end{equation}
where $j\in\{0,1,2\}$, with equal probabilities $\frac{p_0+p_2}{3q'}$ each, and the state $\ket{\text{cat}_1}$, with probability $(1-p_0-p_2)$. The particular value of $q'$ is determined by optimization. These three regions are a reoccurrence of a pattern detected in Ref. \cite{rogers2024quantifying} for mixtures of three neighboring Fock states. Fig. \ref{fig:threeMixedCats} of this paper is comparable to Fig. 3 of Ref. \cite{rogers2024quantifying}. 

\end{widetext}

%

\end{document}